\documentclass[12pt]{iopart}

\usepackage{iopams}  
\usepackage{graphicx}% Include figure files
\expandafter\let\csname equation*\endcsname\relax
\expandafter\let\csname endequation*\endcsname\relax
\usepackage{amsmath}

\usepackage{bm}
\usepackage{hyperref}
\hypersetup{colorlinks=true, allcolors=blue}

\usepackage{upgreek}% non-italic Greek symbols
\usepackage[capitalise]{cleveref}
\newcommand{\micron}{{\upmu \mathrm{m}}}
\newcommand{\Wcmsqd}{{\mathrm{W}\mathrm{cm}^{-2}}}
\DeclareRobustCommand{\rchi}{{\mathpalette\irchi\relax}}
\newcommand{\irchi}[2]{\raisebox{\depth}{$#1\chi$}} % inner command, used by \rchi

\newcommand{\rc}[1]{\textcolor{black}{#1}}
\newcommand{\mc}[1]{\textcolor{black}{#1}}

\usepackage[sort&compress,numbers]{natbib}

\usepackage[ruled,vlined ]{algorithm2e}
\usepackage{algpseudocode} 

\begin{document}
\title{Collimated $\gamma$-ray emission enabled by efficient direct laser acceleration }

\author{K. Tangtartharakul$^1$, G. Fauvel$^2$, T. Meir$^{3,4}$, F. P. Condamine$^{2,5}$, S. Weber$^2$, I. Pomerantz$^3$, M. Manuel$^6$ and A. Arefiev$^1$}

\address{$^1$ Department of Mechanical and Aerospace Engineering, University of California San Diego, La Jolla, CA 92093, USA}
% \address{$^2$ ELI Beamlines Facility, Extreme Light Infrastructure ERIC, Za Radnicí 835, 25241 Dolní Břežany, Czech Republic}
\address{$^2$ The Extreme Light Infrastructure ERIC, ELI Beamlines Facility, Za Radnici 835, 252 41 Dolni Brezany, Czech Republic}
\address{$^3$ The School of Physics and Astronomy, Tel Aviv University, Tel Aviv, 69978, Israel}
\address{$^4$ The School of Electrical Engineering, Tel Aviv University, Tel Aviv, 69978, Israel}
\address{$^5$ GenF, 2 Avenue Gay Lussac, 78990 Élancourt, France}
\address{$^6$ General Atomics, San Diego, California 92121, USA}

%\ead{submissions@iop.org}
\vspace{10pt}
\begin{indented}
\item[]\today \newline
Corresponding Author: Kavin Tangtartharakul \newline
Email: ktangtar@ucsd.edu
\end{indented}

%%%%%%%%%%%%%%%%%%%%%%%%%%%%%%%%%%%%%%%%%%%%%%%%%%%%%%%%%%%%%%%%%%%%%%%%%%%%%%%%%%%%%%%%%%%%%%%%%%%%%%%%%%
\begin{abstract}
\rc{We investigate the mechanisms responsible for single-lobed versus double-lobed angular distributions of emitted \(\gamma\)-rays in laser-irradiated plasmas, focusing on how direct laser acceleration (DLA) shapes the emission profile. Using test-particle calculations, we show that the efficiency of DLA plays a central role. In the inefficient DLA regime, electrons rapidly gain and lose energy within a single laser cycle, resulting in a double-lobed emission profile heavily influenced by laser fields. In contrast, in the efficient DLA regime, electrons steadily accumulate energy over multiple laser cycles, achieving much higher energies and emitting orders of magnitude more energy. This emission is intensely collimated and results in single-lobed profiles dominated by quasi-static azimuthal magnetic fields in the plasma. Particle-in-cell simulations demonstrate that lower-density targets create favorable conditions for some electrons to enter the efficient DLA regime. These electrons can dominate the emission, transforming the overall profile from double-lobed to single-lobed, even though inefficient DLA electrons remain present. These findings provide valuable insights for optimizing laser-driven \(\gamma\)-ray sources for applications requiring high-intensity, well-collimated beams.}

\end{abstract}

%In this paper, we investigate the conditions under which direct laser acceleration (DLA) of electrons in a laser-irradiated plasma can produce distinct photon emission profiles, focusing on the mechanisms responsible for single-lobed versus double-lobed angular distributions of emitted \(\gamma\)-rays. Through a combination of particle-in-cell simulations, test-electron simulations, and theoretical analysis, we demonstrate that the efficiency of DLA is a key determinant of the resulting emission pattern. Our results show that inefficient DLA, characterized by electrons rapidly gaining and losing energy within a single laser cycle, leads to a double-lobed emission profile heavily influenced by laser fields. In contrast, in the efficient DLA regime, where electrons steadily accumulate energy over multiple cycles, the emission is primarily governed by the quasi-static azimuthal magnetic fields generated by the laser in the plasma, resulting in a well-collimated single-lobed emission profile. Additionally, we identify that reducing the electron density in the target enhances the efficiency of DLA, thereby transforming the emission from a double-lobed to a single-lobed profile. These findings provide valuable insights into the optimization of laser-driven \(\gamma\)-ray sources for applications requiring high-intensity, well-collimated beams.

\maketitle
%%%%%%%%%%%%%%%%%%%%%%%%%%%%%%%%%%%%%%%%%%%%%%%%%%%%%%%%%%%%%%%%%%%%%%%%%%%%%%%%%%%%%%%%%%%%%%%%%%%%%%%%%%

\section{Introduction}

The rapid development of lasers and laser technology over the past few decades has paved the path to a variety of novel applications. One such application is the creation of directed and compact laser-driven $\gamma$-ray sources. The construction of high-power laser facilities~\cite{Danson.HPLSE.2019.Worldwide}, including ELI Beamlines~\cite{Borneis.HPLSE.2021.ELIbeamlines}, ELI-NP~\cite{Doria.JI.2020.ELINP}, Apollon~\cite{Cheriaux.AIP.2017.APOLLON}, and CoReLS~\cite{Yoon.Optica.2019.CoReLS}, has motivated the scientific community to examine the role of ultra-high-laser intensities. Based on computational and theoretical research, there is a general consensus that the efficiency of generating $\gamma$ -rays --- a key parameter for $\gamma$-ray sources --- can be greatly enhanced by increasing the peak laser intensity. Specifically, a significant increase in efficiency is expected as the peak intensity rises from $10^{21}~\Wcmsqd$ to $10^{23}~\Wcmsqd$. The newly constructed laser facilities are anticipated to offer intensities exceeding $10^{22}~\Wcmsqd$, thereby opening new regimes of light-matter interaction for exploration. For the first time, these regimes may allow for the conversion of more than several percent of laser energy into \mc{directed} $\gamma$-rays~\cite{nakamura.prl.2012, Ji.PRL.2014.Radiation, Nerush.PoP.2014.Gamma, Stark.PRL.2016.Emission, Lezhnin.PoP.2018.High, Wang.PRA.2020.Pair,  Hadjisolomou.SR.2022.Towards, He.CP.2021.Dominance, Vladisavlevici.PPCF.2024.Investigation}.

The energy conversion efficiency is not the only important parameter of laser-driven $\gamma$-ray sources, but it is the one that has received particular attention. This is because experimentally demonstrated laser-driven sources convert the incident laser energy into x-rays and $\gamma$-rays with low efficiency. For example, a source employing laser-driven wakefield acceleration of electrons produces $10^9$ photons with energies above 1~keV using an 11~J laser beam~\cite{Cole.SR.2015.Wakefield}. The peak of the photon spectrum is in the hard x-ray range with $\varepsilon_{\gamma} \approx 30$~keV. Using this value, we estimate that the energy conversion efficiency is less than $10^{-6}$. Such a source has been successfully used for micro-computed tomography~\cite{Cole.SR.2015.Wakefield}. However, the low conversion efficiency does limit the absolute number and density of emitted photons particularly in the $\gamma$-ray range, precluding the use of the existing sources for certain studies and applications. Two notable examples include radiography of high-Z materials~\cite{Espy.RSI.2021.HighZ,Wang.NIMPRSA.2023.Radiography} and the creation of electron-positron pairs from light via photon-photon collision~\cite{Ribeyre.PRE.2016.Pair,Ribeyre.PPCF.2016.Breit,Ribeyre.PPCF.2018.Breit,Wang.PRA.2020.Pair,He.CP.2021.Dominance, Sugimoto.PRL.2023.Positron}.

The desire to drastically increase the $\gamma$-ray number and density has driven the search for novel regimes of laser-matter interactions. An important consideration is that the emission of $\gamma$-rays in laser-driven $\gamma$-ray sources is not a direct conversion of optical photons into $\gamma$-rays. Instead, the laser first transfers its energy to plasma electrons in the laser-irradiated material. These energized electrons then emit the $\gamma$-rays. This two-step process highlights the crucial role of plasma electrons in mediating the energy transfer from the laser to the emitted $\gamma$-rays. It also suggests that substantive efficiency improvements must involve changes in electron physics that can amount to switching from the laser-driven wakefield acceleration to another mechanism.

A promising and qualitatively different regime of electron acceleration can be accessed using an ultra-high-intensity high-power laser and a dense plasma. A key effect in this scenario is relativistically induced transparency~\cite{Gibbon.WS.2005.Short, Kaw.PhysicsFluids.1970.RT, Palaniyappan.NaturePhysics.2012.RT}. Essentially, the laser heats the electrons in the dense plasma to relativistic energies, increasing their effective mass and making the plasma transparent. As the laser propagates through the plasma, it directly accelerates electrons to ultra-relativistic energies while simultaneously driving strong plasma magnetic~\cite{Stark.PRL.2016.Emission, Gong.PRE.2020.Magnetic} and electric fields~\cite{Pukhov.PoP.1999.Channel, Arefiev.PoP.2016.Beyond,Cohen.SA.2024.Undepleted}. These electrons interact with the laser and plasma fields, emitting energetic photons.

As demonstrated in multiple publications~\cite{nakamura.prl.2012,Stark.PRL.2016.Emission, Lezhnin.PoP.2018.High, Wang.PRA.2020.Pair, Meir.PRA.2024.Compton}, this regime offers a path to enhancing energy conversion efficiency by increasing the number of energetic electrons and the fields that induce their emission. Both the laser amplitude and the plasma density are crucial. High laser intensity is essential for achieving ultra-relativistic electron energies through direct laser acceleration (DLA). Meanwhile, high plasma density is important for two reasons: 1) it allows for the acceleration of a large number of electrons, and 2) it enables the generation of strong plasma fields that are vital for photon emission.

The predicted increase in efficiency is dramatic: $\gamma$-ray sources based on DLA are expected to convert several to tens of percent of laser energy into $\gamma$-ray~\cite{nakamura.prl.2012,Stark.PRL.2016.Emission, Lezhnin.PoP.2018.High, Wang.PRA.2020.Pair}. However, this increase comes at a cost. The resulting beam typically lacks strong collimation and often exhibits a two-lobed structure~\cite{Huang.PRE.2016.Characteristics,Wang.PoP.2015.Gamma,Galbiati.FP.2023.Numerical,Xue.MRE.2020.Generation,Rinderknecht.NJP.2021.Relativistically,Jansen.PPCF.2018.Pair}, which has become associated with this regime. So far, there has been relatively little effort aimed at improving the collimation or the angular pattern to achieve a well-directed beam of $\gamma$-rays without multiple lobes. This might be due to the fact that the mechanism that determines the angular distribution is not well understood. Nonetheless, enhancing beam collimation is of paramount significance in unlocking the practical utility of laser-driven $\gamma$-ray beams.

\mc{In this paper, we systematically investigate the conditions under which DLA of electrons results in distinct photon emission profiles, focusing on the transition between single-lobed and double-lobed angular distributions of emitted $\gamma$-rays. By combining particle-in-cell (PIC) simulations, test-electron simulations, and theoretical analysis, we identify the efficiency of DLA as a crucial factor in determining the emission pattern. Specifically, we demonstrate that inefficient DLA — where electrons rapidly gain and lose energy within a single laser cycle — results in a double-lobed emission profile heavily influenced by the laser fields. Conversely, in the efficient DLA regime, where electrons steadily accumulate energy over multiple cycles, the emission is primarily governed by the quasi-static azimuthal magnetic fields in the plasma, producing a well-collimated, single-lobed emission profile. Furthermore, electrons in the efficient DLA regime achieve much higher energies and, as a result, emit many orders of magnitude more energy. This allows them to dominate the emission and transform the overall plasma emission profile from double-lobed to single-lobed, even though inefficient DLA electrons may still be present. Using PIC simulations, we also show that reducing the electron density in the target promotes efficient DLA, thereby shifting the emission from a double-lobed to a single-lobed profile. These findings have significant implications for optimizing laser-driven $\gamma$-ray sources for practical applications. }

%In contrast, in the efficient DLA regime, electrons steadily accumulate energy over multiple laser cycles, achieving much higher energies and emitting many orders of magnitude more energy. This emission is intensely collimated and results in single-lobed profiles dominated by quasi-static azimuthal magnetic fields in the plasma. Particle-in-cell simulations demonstrate that lower-density targets create favorable conditions for some electrons to enter the efficient DLA regime. These electrons can dominate the emission, transforming the overall profile from double-lobed to single-lobed, even though inefficient DLA electrons remain present. These findings provide valuable insights for optimizing laser-driven \(\gamma\)-ray sources for applications requiring high-intensity, well-collimated beams.

The rest of the paper is structured as follows: In \cref{sec: pic motivation}, we present two 3D particle-in-cell (PIC) simulations of laser-driven $\gamma$-ray emission, showing that one can switch from a conventional two-lobed emission profile to a collimated single-lobed profile by lowering the target density. \mc{Using the PIC results as a motivation,} \cref{sec: simple estimates} explores two limiting cases of electron motion: an electron interacting solely with the magnetic field of a filament and an electron irradiated by only a plane electromagnetic wave. This analysis provides foundational insights into the angular distribution of emitted photons. Specifically, we demonstrate that the plasma magnetic filament induces a single-lobe emission profile. In \cref{sec: test electron model}, we use a test-electron model to analyze the combined effects of laser and plasma fields on electron dynamics, distinguishing between efficient and inefficient direct laser acceleration (DLA) regimes. We find that the efficient regime has strong parallels to the magnetic filament limiting case considered in \cref{sec: simple estimates}. Namely, the emission process is dominated by the magnetic filament and results in a single-lobed angular distribution in the photon emission despite an electron angular distribution with two lobes. Next, in \cref{sec: parameter scan}, we perform a parameter scan using the test-electron model to demonstrate that the findings in \cref{sec: test electron model} apply broadly across the parameter space. Finally, \cref{Sec: summary and discussion} offers a summary of our findings, discussing their implications for optimizing laser-driven $\gamma$-ray sources.

%Next, in \cref{sec: parameter scan}, we perform a parameter scan on our test-electron model to show that our findings in \cref{sec: test electron model} apply broadly across the parameter space. 

\begin{figure*}
    \centering
    \includegraphics[width=1\textwidth]{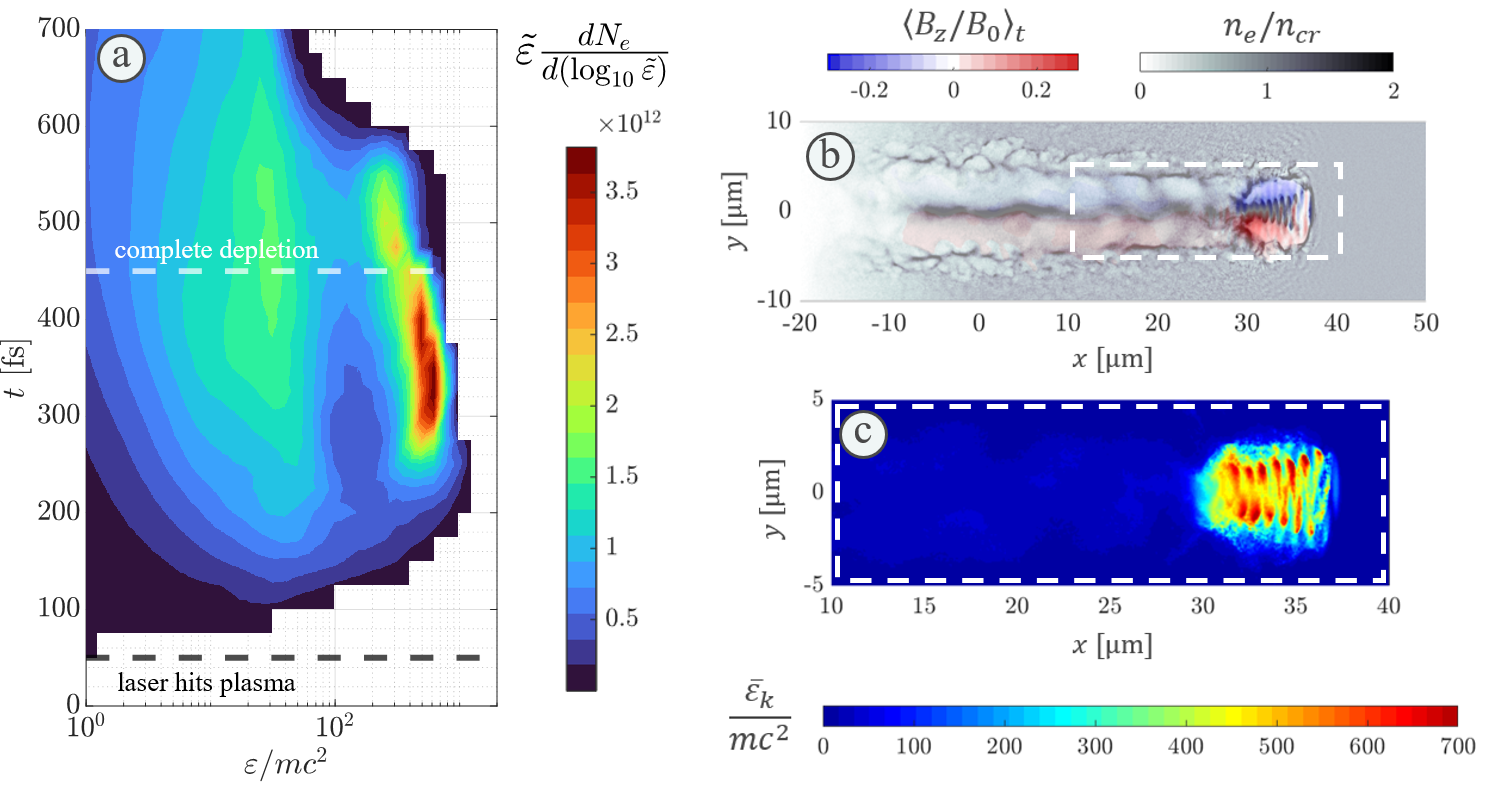}
    \caption{3D PIC simulation results for the target with $n_e = 1 n_{cr}$ irradiated by the laser described in \cref{table: pic}.  
    (a)~Time evolution of the energy-weighted log-energy electron distribution function, as defined by \cref{def F}.  (b)~Time-averaged magnetic field, $\langle B_z / B_0 \rangle_t$, and instantaneous electron density, $n_e/n_{cr}$, at $t=300$ fs. The time-averaging is performed over five laser periods. (c)~Cell-averaged electron kinetic energy at $t=300$ fs within the zoomed-in region marked by the dashed-bordered box in panel (b). }
    \label{fig: dla background}
\end{figure*}

%%%%%%%%%%%%%%%%%%%%%%%%%%%%%%%%%%%%%%%%%%%%%%%%%%%%%%%%%%%%%%%%%%%%%%%%%%%%%%%%%%%%%%%%%%%%%%%%%%%%%%%%%%

\section{Density reduction as a path to collimated $\gamma$-ray emission in 3D PIC simulations} \label{sec: pic motivation}

In this section, we present two three-dimensional (3D) particle-in-cell (PIC) simulations of laser-driven \(\gamma\)-ray emission: one resulting in a conventional two-lobed emission profile and the other producing a collimated profile with a single peak. Our detailed analysis of the laser-driven electron acceleration and subsequent photon emission reveals that enhanced electron energy gain is the crucial factor in achieving the single-peak profile.
 
In our PIC simulations, we use the same laser beam, varying only the target density to demonstrate its effectiveness as a control parameter. To directly link with current experimental capabilities, we selected laser parameters similar to those of the ELI-Beamline's L3 laser system~\cite{Borneis.HPLSE.2021.ELIbeamlines}. \Cref{table: pic} provides the simulation details. In our setup, the laser beam irradiates a flat target tilted at a $20^{\circ}$ angle relative to the laser beam axis. The tilt is an experimental requirement to prevent laser reflection off the target back into the laser optics system.

In both simulations, the target is initialized as a uniform fully ionized carbon plasma. The target electron density for both simulations is specified in \Cref{table: pic}, with values of $n_e=1n_{cr}$ and $n_e=4n_{cr}$, where $n_{cr} = m \omega^2 / 4 \pi e^2$ is the classical critical (cutoff) density for a laser pulse with frequency $\omega \equiv 2 \pi c /\lambda_0$. Here, $m$ is the electron rest mass, $e$ is the electron charge, $c$ is the vacuum speed of light, and $\lambda_0$ is the vacuum wavelength of the laser. Experimentally, such targets can potentially be realized using plastic (CH) foams with low mass-density and sub-wavelength pore size~\cite{Nagai.PoP.2018.Porous, Hund.FST.2006.Aerogel}. The electron density for a fully ionized CH foam target after homogenization is given by $n_e / n_{cr} \approx 0.2 \rho~[\mbox{mg/cm}^{3}]$, where $\rho$ is the mass density of the foam. For reference, an electron density range of $1 \lesssim n_e/n_{cr} \lesssim 10$ corresponds to $5 \mbox{ mg/cm}^{3} \lesssim \rho \lesssim 50 \mbox{ mg/cm}^{3}$.  Achieving $n_e= 1n_{cr}$ is challenging but possible, while foams with the mass density needed to achieve $n_e=4n_{cr}$ are commercially available.

\begin{table} \centering
\footnotesize
\begin{tabular}{ | p{4.5cm} | p{8.7cm}|}
\hline
\multicolumn{2}{|c|}{Laser parameters} \\
\hline
Peak cycle averaged power   & 0.70 PW\\
\hline
Peak intensity & 3.81 $\times 10^{21}~\Wcmsqd$\\
\hline
Total incident energy &  $21$ J\\
\hline
Normalized amplitude,  $a_{0}$ & 42.7\\
\hline
E-field peak amplitude, $E_{0}$ & 1.66 $\times 10^{14}$~V/m \\
\hline
B-field peak amplitude, $B_{0}$ & 5.55 $\times 10^{5}$~T \\
\hline
Vacuum wavelength, $\lambda_{0}$ & $0.81~\micron$\\
\hline
Waist-size, $w_{0}$ &  4.25 $\micron$ (intensity FWHM is 5 $\micron$) \\
\hline
Focal plane's position & $x = 0$ (front of the target)\\
\hline
Propagation & +$\bm{x}$ direction \\
\hline
Polarization & linear ($E_y$ only in the focal plane)\\
\hline
Temporal profile (intensity) & Gaussian (intensity FWHM is 27~fs)\\
\hline
Transverse profile (intensity) & Gaussian with $I(x=0) \propto \exp(-2 r^2/w_0^2)$ \\
\hline
\hline
\multicolumn{2}{|c|}{Plasma and domain parameters} \\
\hline
Composition & Fully ionized carbon\\
\hline
Critical density, $n_{cr}$ &   $1.7 \times 10^{21}$ cm$^{-3}$\\
\hline
Target density, $n_{e}$ & $4n_{cr}$ and $1n_{cr}$\\
\hline
Density ramp (front) & \mc{linearly rises from 0 to $n_e$ over $-20~\micron<x<0~\micron$} \\
\hline
Target tilt & 20 degrees \\
\hline
Target dimensions & $40~\micron \times 15~\micron \times 15~\micron$ and $100~\micron \times 15~\micron \times 15~\micron$  \\
\hline
Spatial resolution & 15 cells per $\micron$ in all three dimensions\\
\hline
Macro-particles & 5 electrons per cell\\
             & 5 carbon ions per cell\\
\hline
\end{tabular}
\caption{Parameters of the two 3D PIC simulations presented in \cref{sec: pic motivation}.}
\label{table: pic}
\end{table}

In both cases, the laser pulse easily propagates into the target due to relativistically induced transparency~\cite{Gibbon.WS.2005.Short}, driving a strong quasi-static azimuthal plasma magnetic field. \Cref{fig: dla background}b illustrates this for the case with $n_e= 1n_{cr}$, where the laser propagation can be inferred from the perturbation of the electron density $n_e/n_{cr}$. It is important to note that the transparency of the classically opaque plasmas in our study is enabled by the high laser amplitude. The specific criterion for this is \(a_0 \gg 1\), where \(a_0 \equiv |e| E_0 / m c \omega\) represents the normalized laser amplitude, with \(E_0\) being the peak electric field amplitude. At $a_0 \gtrsim 1$, the laser fields induce relativistic electron motion, causing the cutoff density to become dependent on the laser amplitude. As a result, a plasma with $n_e < a_0 n_{cr}$ becomes transparent to the laser pulse, even if the plasma is classically overdense ($n_e \geq n_{cr}$)~\cite{Gibbon.WS.2005.Short}.

In the regime of laser propagation enabled by relativistically induced transparency, plasma electrons can gain energies that far exceed the typical oscillatory energy, estimated as $a_0 m c^2$~\cite{Pukhov.PoP.1999.Channel,Arefiev.PoP.2016.Beyond,Gong.PRE.2020.Magnetic}. \Cref{fig: dla background}a illustrates this for the case with $n_e= 1n_{cr}$ by showing the time evolution of an energy-weighted log-energy electron distribution function across the entire simulation domain. The displayed quantity is $\tilde{\varepsilon} F$, where $F$ is a log-energy electron distribution function and  $\tilde{\varepsilon} = \varepsilon/m c^2$ is the electron kinetic energy normalized to $mc^2$. We compute $F$ using the conventional distribution function $f = dN_e/d \varepsilon$ outputted by the code, according to the relation
\begin{equation} \label{def F}
 F \equiv \frac{dN_e}{d \log_{10} \tilde{\varepsilon}} = f(\varepsilon) \ln(10) \varepsilon .
\end{equation}
It is evident from \cref{fig: dla background}a that a significant fraction of the energy transferred from the laser to the electrons is carried by electrons with $\varepsilon \approx 400 mc^2 \approx 200$~MeV (e.g. see the horizontal slice at $t = 300$~fs), which is an order of magnitude higher than $a_0 m c^2 \approx 22$~MeV. The energy is transferred directly from the laser to electrons, which is the reason why the underlying mechanism is referred to as the direct laser acceleration (DLA). The energy comes from the transverse laser electric field that drives transverse electron oscillations. These oscillations are clearly seen in \cref{fig: dla background}c, which shows the cell-averaged electron kinetic energy. At $a_0 \gg 1$, the laser magnetic field redirects the accumulated energy of relativistic electrons towards the longitudinal motion with the help of the $\bm{v} \times \bm{B}$ force. \mc{Trajectories of electrons experiencing DLA in a setup with a structured target are available in Ref.~\cite{Gong.PRE.2020.Magnetic}.}

The significant energy gain observed in \cref{fig: dla background}a is facilitated by the azimuthal plasma magnetic field shown in \cref{fig: dla background}b. As the laser pulse travels through the plasma, it induces a magnetic filament configuration, maintained by a co-axial electron current structure. This quasi-static magnetic field deflects outward-moving electrons back towards the laser beam axis~\cite{Wang.PoP.2020.Magnetic}, effectively providing transverse electron confinement. Throughout the paper, we will refer to this configuration as a magnetic filament. When the frequency of these transverse deflections aligns with the laser field oscillations at the electron's location, it can lead to enhanced energy gain via betatron resonance~\cite{Pukhov.PoP.1999.Channel,Khudik.PoP.2016.Universal}. This mechanism is explored in depth in \cref{sec: test electron model}, but it is crucial to note here that this energy gain process is fundamentally different from the acceleration of an electron in a vacuum~\cite{Arefiev.PoP.2016.Beyond}.

The laser and plasma fields acting on plasma electrons induce strong acceleration, which causes them to emit photons. In what follows, the subscript $\gamma$ denotes quantities related to photons. The emission by a moving electron as it undergoes acceleration due to macroscopic electric and magnetic fields ($\bm{E}$ and $\bm{B}$) is characterized by a single dimensionless parameter~\cite{Jackson.Wiley.1999.Classical}
\begin{equation} \label{chi}
    \rchi= \frac{\gamma}{B_{crit}} \left| \bi{E} - \frac{\bi{p} (\bi{p} \cdot \bi{E})}{p^2} + \frac{1}{\gamma m c} \left[ \bi{p} \times \bi{B} \right] \right|,
\end{equation}
where $\bm{p}$ is the electron momentum,
\begin{equation}
    \gamma = \sqrt{1 + p^2/ m^2 c^2}
\end{equation}
is its relativistic factor, and $B_{crit}=m^2 c^3 / |e| \hbar \approx 4.4 \times 10^{13}$~G is the critical magnetic field defined using the reduced Planck constant $\hbar$. The power emitted by the electron~\cite{Ridgers.JCP.2014.Modelling, Kirk.PPCF.2009.Pair} is
\begin{equation}
    P_{\gamma} =  \frac{2 \alpha_f}{3}  \frac{(m c^2)^2}{\hbar}    \rchi^2,  \label{eq: gamma power}
\end{equation}
where $\alpha_f = e^2/ \hbar c$ is the fine-structure constant. We assume that $\rchi \ll 1$, which holds true in our simulations. Consequently, the gaunt factor~\cite{Ridgers.JCP.2014.Modelling, Kirk.PPCF.2009.Pair}, the quantum correction factor to our classical description, is omitted. The spectrum of emitted photons has a peak~\cite{Jackson.Wiley.1999.Classical} at the energy 
\begin{equation}
    \varepsilon_{\gamma} \approx 0.4 \rchi  \gamma m c^2. \label{eq:ph_energy}
\end{equation}

\begin{figure} [b!]
        \centering
        \includegraphics[width=0.85\linewidth]{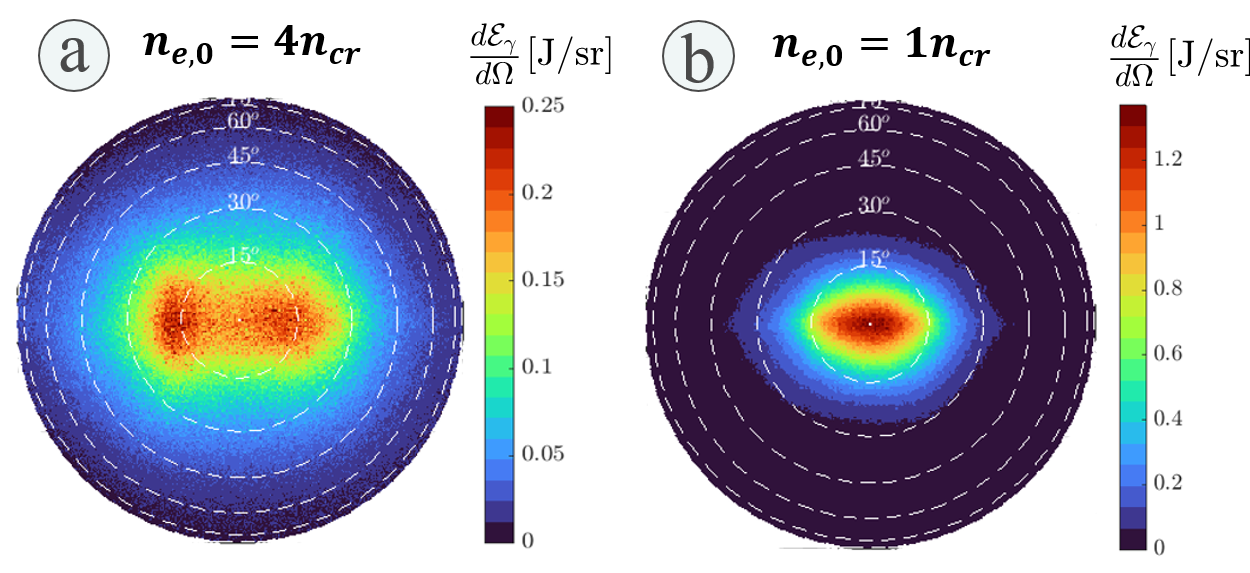}
    \caption{Angular distribution of the emitted energy, $d {\cal{E}}_{\gamma} / d \Omega$, from simulations with $n_{e,0} = 4n_{cr}$ (a) and $n_{e,0} = 1n_{cr}$ (b). Only photons with $\varepsilon_{\gamma}> 10$~keV are included in the plots. The dashed circles indicate polar angles in $15^{\circ}$ increments, with the laser beam direction corresponding to $0^{\circ}$.}
    \label{fig: pic photons}
\end{figure}

We primarily focus on emission by ultrarelativistic electrons (\(\gamma \mspace{-5mu} \gg \mspace{-5mu} 1\)), as these are the electrons capable of emitting hard x-rays and \(\gamma\)-rays according to \cref{eq:ph_energy}. Ultrarelativistic electrons emit primarily along their momentum \(\bm{p}\), with the emitted power concentrated within a narrow cone with an opening angle of \(1/\gamma\). In our simulations, we neglect the angular spread, so the photons are emitted as individual particles in the direction of the electron's momentum or velocity. The emission is calculated during the PIC simulation using a Monte Carlo algorithm that probabilistically determines the emitted photon energy based on the value of \(\rchi\)~\cite{Ridgers.JCP.2014.Modelling, Gonoskov.PRE.2015.Schemes}. Following  emission, the electron experiences a recoil equal to the emitted photon's momentum.

\begin{figure}
    \centering
    \includegraphics[width=0.8\linewidth]{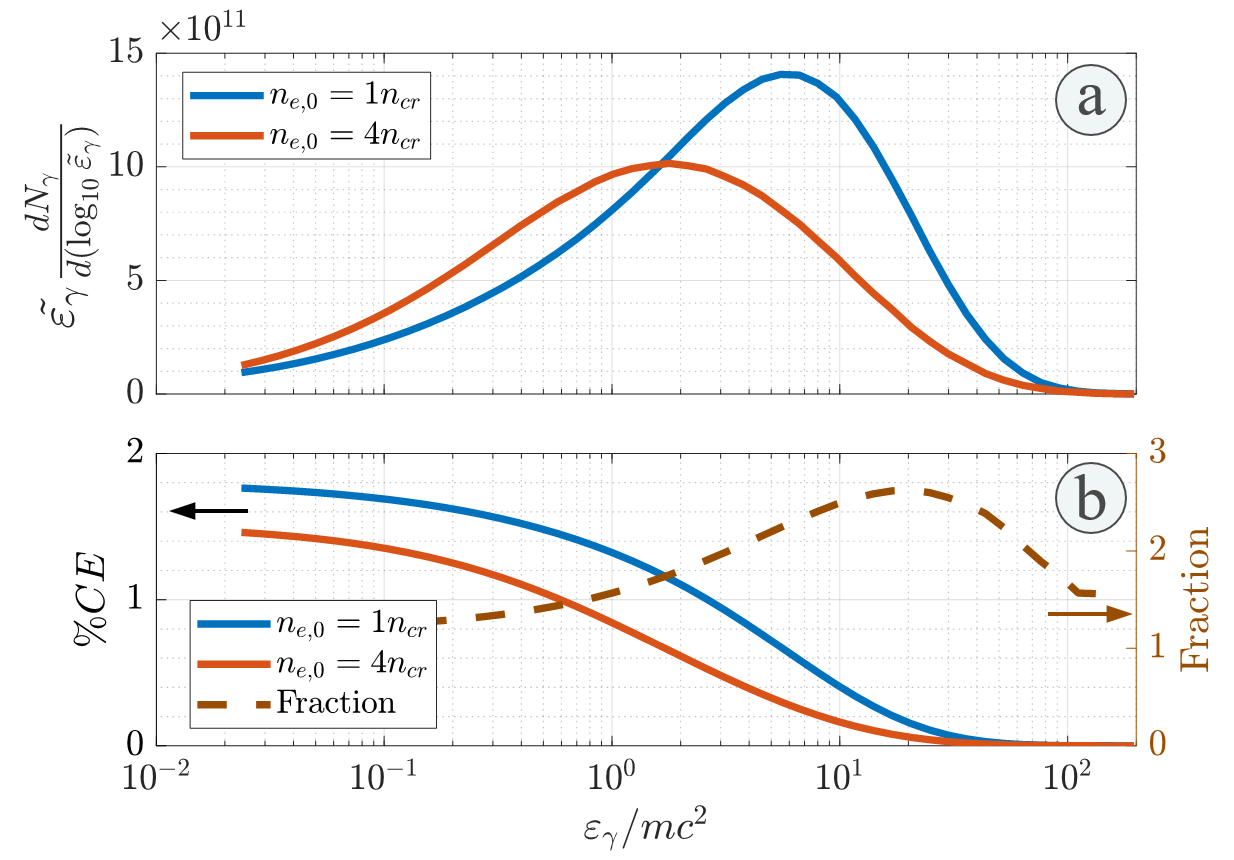}
    \caption{Spectra of emitted photons and conversion efficiency of laser energy into photons for simulations with $n_{e,0} = 1 n_{cr}$ and $n_{e,0} = 4 n_{cr}$. (a) Energy-weighted log-energy distribution functions of all photons emitted during each simulation, where $\tilde{\varepsilon}_{\gamma} \equiv \varepsilon_{\gamma} / m c^2$. (b) Conversion efficiencies of laser energy into photons for each simulation. The conversion efficiency ($\% CE$) represents the percentage of the total laser energy that is emitted as photons whose individual energy is above $\varepsilon_{\gamma}$.  The dashed curve shows the ratio of conversion efficiencies between the two simulations (using the $y$-axis on the right).  }
    \label{fig: photon distw CE}
\end{figure}

\begin{figure}
    \centering
    \includegraphics[width=0.8\linewidth]{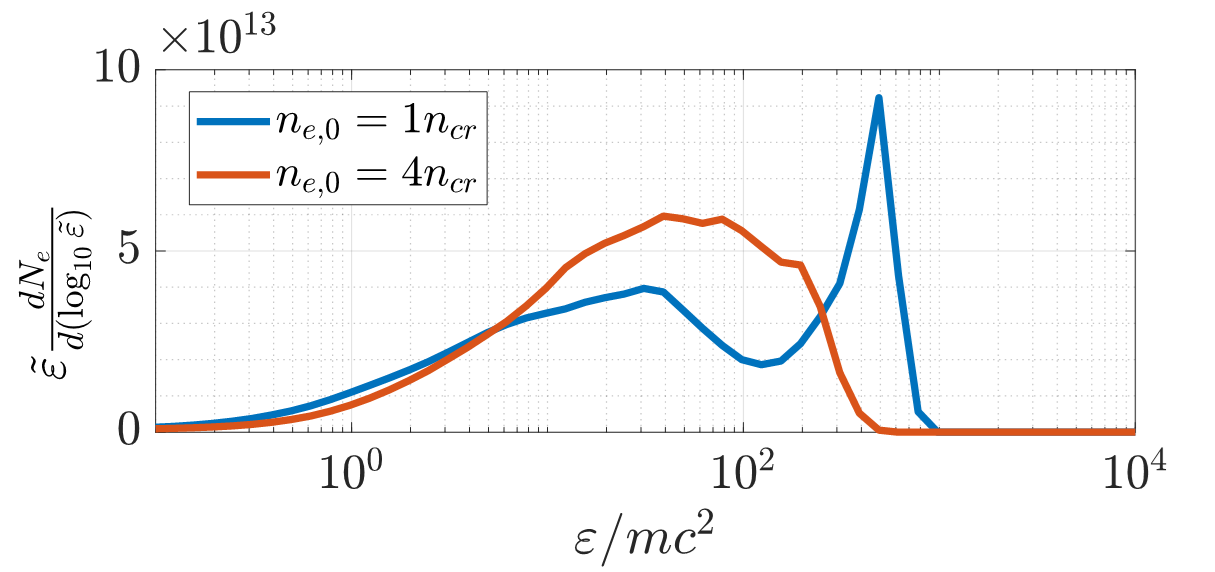}
    \caption{Snapshots of energy-weighted log-energy electron distribution functions for simulations with $n_{e,0} = 1 n_{cr}$ (blue) and $n_{e,0} = 4 n_{cr}$ (red). The snapshots are taken at the moments when the total kinetic energy of the electron population reaches its peak value ($t=400$~fs and $t=225$~fs, respectively). The distribution function is defined by \cref{def F}.  
    }
    \label{fig: electron distw}
\end{figure}

The photon emission patterns in \cref{fig: pic photons} from the two simulations exhibit a strong angular dependence on the initial target electron density $n_{e,0}$. The depicted quantity is the accumulated emitted energy per steradian, obtained by projecting photons with $\varepsilon_{\gamma} > 10$~keV onto a sphere. The polar angle is measured with respect to the propagation direction of the laser pulse, with the dashed circles marking polar angles in $15^{\circ}$ increments. The horizontal plane represents the polarization plane of the laser electric field. In the simulation with $n_{e,0} = 4n_{cr}$ (\cref{fig: pic photons}a), the emission pattern displays a two-lobed structure in the laser polarization plane. In contrast, the simulation with $n_{e,0} = 1n_{cr}$ (\cref{fig: pic photons}b) shows an emission pattern with a single lobe aligned with the direction of the laser propagation. \mc{\ref{Appendix A} provides additional plots for three different photon energy ranges, demonstrating that the observed angular dependence persists across different photon energies and is not specific to the chosen energy threshold of $10$~keV.}

The switch from two lobes to one results in a significant increase in the emitted energy per steradian. Upon integrating over the entire sphere, we found that both targets convert roughly 2\% of the incident laser energy into photons with $\varepsilon_{\gamma} > 10$~keV, emitting approximately the same amount of energy: 0.31~J for $n_{e,0} = 4n_{cr}$ and 0.37~J for $n_{e,0} = 1n_{cr}$. However, the angular distribution for $n_{e,0} = 1n_{cr}$ is narrower, partly because it contains only a single lobe. As a result, the peak energy per steradian is approximately five times higher than in the simulation with $n_{e,0} = 4n_{cr}$. This aspect, combined with the fact that the direction of the single lobe aligns with the known direction of the laser beam, makes the emission regime at $n_{e,0} = 1n_{cr}$ more attractive for applications.

\Cref{fig: photon distw CE} compares the spectra of emitted photons and the conversion efficiencies of laser energy into photons for the two simulations with different target densities. \Cref{fig: photon distw CE}a presents energy-weighted log-energy photon distribution functions. In both cases, a peak is present, but it occurs at a higher photon energy for $n_{e,0} = 1 n_{cr}$ ($6 mc^2 \approx 3$~MeV) compared to $n_{e,0} = 4 n_{cr}$ ($1.5mc^2 \approx 0.8$~MeV). \Cref{fig: photon distw CE}b illustrates the conversion of laser energy into photons with energies above $\varepsilon_{\gamma}$ for both simulations. The dashed curve represents the ratio of the two conversion efficiencies. Notably, the target with $n_{e,0} = 1 n_{cr}$ becomes significantly more efficient than the target with $n_{e,0} = 4 n_{cr}$ at $\varepsilon_{\gamma} \gtrsim 3mc^2 \approx 1.5$~MeV. 

The differences in photon spectra and conversion efficiency shown in \cref{fig: photon distw CE} suggest that plasma electrons experience more efficient DLA in the simulation with \(n_{e,0} = 1 n_{cr}\). To validate this, we compared the electron spectra from both simulations. \Cref{fig: electron distw} presents snapshots taken at the times when the total kinetic energy of the electron population reaches its maximum — $t=400$~fs for the \(n_{e,0} = 1 n_{cr}\) simulation and $t=225$~fs for the \(n_{e,0} = 4 n_{cr}\) simulation. As expected, the peak in \cref{fig: electron distw} shifts to higher electron energies in the simulation with \(n_{e,0} = 1 n_{cr}\). By reaching higher \(\gamma\) values, these electrons also achieve higher \(\chi\) values, which, though not shown, we confirmed from the simulations. This accounts for why electrons in the \(n_{e,0} = 1 n_{cr}\) simulation emit higher energy photons [see \cref{eq:ph_energy}] with greater power [see \cref{eq: gamma power}].

The key question is why the more efficient DLA at \(n_{e,0} = 1 n_{cr}\) results in photon emission that is better collimated and exhibits only a single lobe in the angular distribution. To gain further insight into the emission process, we tracked \(0.01\%\) of the electron population in each simulation (192,972 and 96,936 macro-particles in the simulations with \(n_{e,0} = 1 n_{cr}\) and \(n_{e,0} = 4 n_{cr}\), respectively), recording the emission information for each electron. From the tracked population, we selected the highest-energy electron for each run, as the emission process favors higher \(\gamma\). In the simulation with \(n_{e,0} = 4 n_{cr}\), the highest-energy electron achieved \(\gamma \approx 700\), whereas in the simulation with \(n_{e,0} = 1 n_{cr}\), the highest-energy electron achieved \(\gamma \approx 1200\). Since the difference between the double and single lobe distributions is particularly pronounced in the laser-polarization plane, we examine the electron dynamics in the \((x,y)\)-plane. The electrons emit along \(\bm{p}\), so the polar angle \(\theta_{xy} \equiv \tan^{-1}(p_y/p_x)\) is directly linked to the angular distribution of the photon emission.

\begin{figure} [t!]
        \centering
        \includegraphics[width=0.7\linewidth]{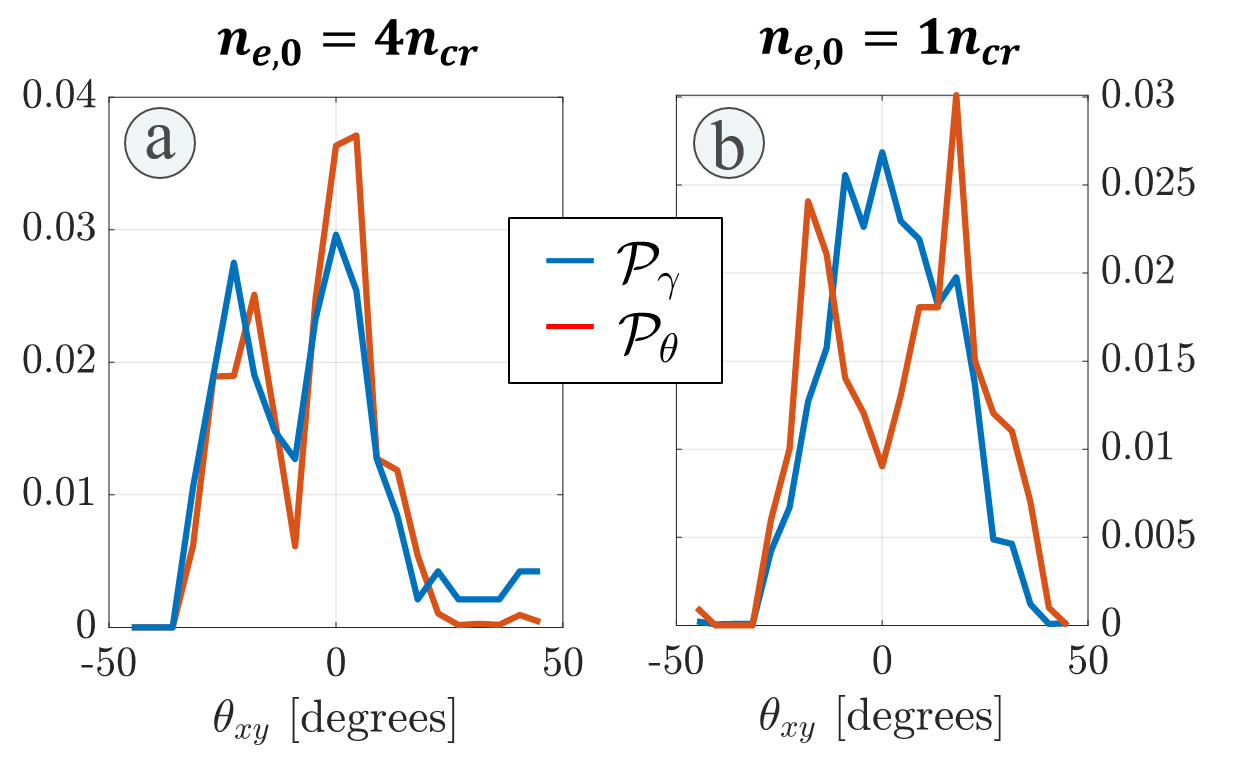}
    \caption{Angular probability $\mathcal{P_{\theta}}$ and angular distribution of emitted energy $\mathcal{P}_{\gamma}$ for the two tracked electrons. The polar angle is defined as $\theta_{xy} \equiv \tan^{-1} (p_y/p_x)$. Panel (a) corresponds to $n_{e,0} = 4n_{cr}$, and panel (b) corresponds to $n_{e,0} = 1n_{cr}$. The definitions of $\mathcal{P_{\theta}}$ and  $\mathcal{P}_{\gamma}$, along with the electron selection criterion, are given in the text.} \label{fig: dist pic}
\end{figure}

\Cref{fig: dist pic} compares the fraction of time each electron spends with $\bm{p}$ at $\theta_{xy}$ and the amount of energy radiated in that direction. \Cref{fig: dist pic}a corresponds to the electron in the simulation with $n_{e,0} = 4n_{cr}$, while \Cref{fig: dist pic}b corresponds to the electron in the simulation with $n_{e,0} = 1n_{cr}$. The solid red curve represents the angular probability distribution $\mathcal{P}_{\theta}$, where the integral $\int_{\theta_1}^{\theta_2} \mathcal{P}_{\theta} d \theta$ gives the probability that the electron’s momentum in the $(x,y)$-plane has an angle between $\theta_1$ and $\theta_2$ over the entire trajectory. The solid blue curve represents the angular distribution of the emitted power $\mathcal{P}_{\gamma}$, where $\int_{\theta_1}^{\theta_2} \mathcal{P}_{\gamma} d \theta$ is the fraction of emitted energy in the direction $\theta_1 \leq \theta_{xy} \leq \theta_2$. We observe that both electrons spend most of their time at angles away from $0^{\circ}$, producing the two-lobed profile of $\mathcal{P}_{\theta}$. However, the emitted power distribution differs qualitatively between the two electrons: in the simulation with $n_{e,0} = 4n_{cr}$, the electron produces two peaks, but in the simulation with $n_{e,0} = 1n_{cr}$ the electron produces only a single peak. \mc{Although these results are shown for two randomly selected electrons, they are representative of the energetic electron populations in the respective simulations. The changes in the angular distribution of the emitted power match the qualitative changes seen in \cref{fig: pic photons}, showing the distribution for the entire electron population.} 

The results shown in \cref{fig: dist pic} lead to an important observation. We infer that in the simulation with $n_{e,0} = 1n_{cr}$, the electron dynamics causes the electron to disproportionately emit at small angles of the trajectory, even though the electron spends more time at larger angles. This observation is not limited to the two electrons we examined; a similar trend is observed for other tracked electrons.

%****************************

\section{Emission analysis: two limiting cases} \label{sec: simple estimates}

\begin{figure}[b]
    \centering
    \includegraphics[width=1.0\linewidth]{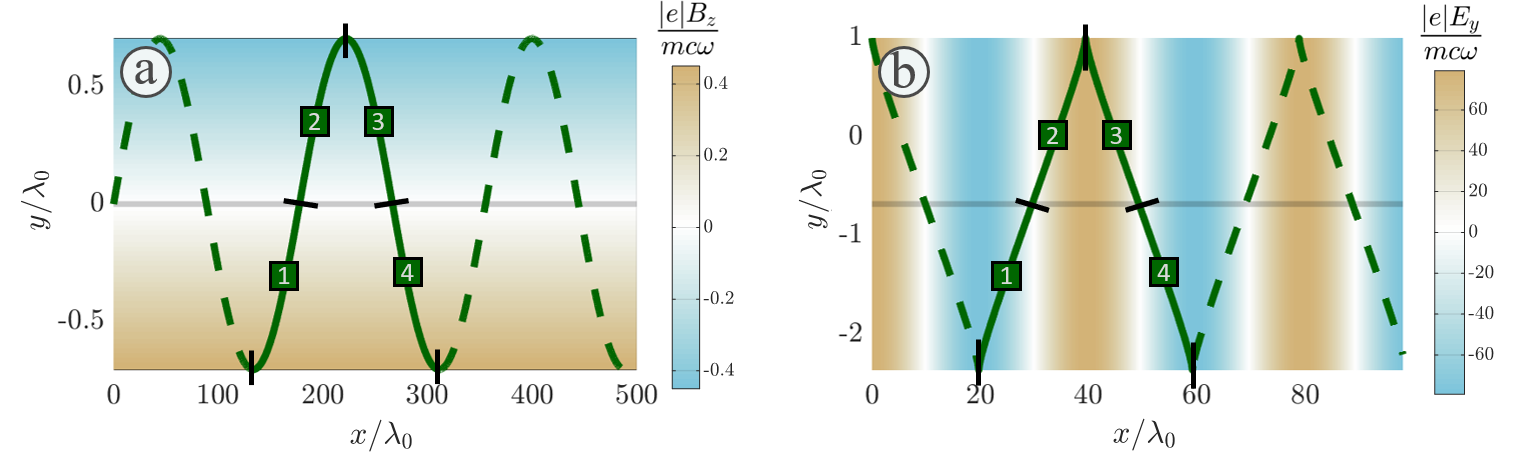}
    \caption{Electron trajectories for an electron (a) moving along a magnetic field filament without laser fields and (b) irradiated only by a plane electromagnetic wave in a vacuum. The parameters used to generate the trajectories are given in the caption for \cref{fig: DLAfilamanet_dist_MATLAB}. In both cases, the solid curve represents one full oscillation, with the numbers marking four distinct quarters. In panel (a), the quarters correspond to: (1) $\theta > 0$ and $d \theta / dt > 0$; (2) $\theta > 0$ and $d \theta / dt < 0$; (3) $\theta < 0$ and $d \theta / dt < 0$, and (4) $\theta < 0$ and $d \theta / dt > 0$. In panel (b), they correspond to: (1) $\pi/2 \geq \theta \geq \theta_{\min}$ and $d \theta / dt < 0$; (2) $\pi/2 \geq \theta \geq \theta_{\min}$ and $d \theta / dt > 0$; (3) $-\pi/2 \leq \theta \leq -\theta_{\min}$ and $d \theta / dt > 0$, and (4) $-\pi/2 \leq \theta \leq -\theta_{\min}$ and $d \theta / dt < 0$.}
    \label{fig:traj}
\end{figure}

\begin{figure}
    \centering
    \includegraphics[width=0.8\linewidth]{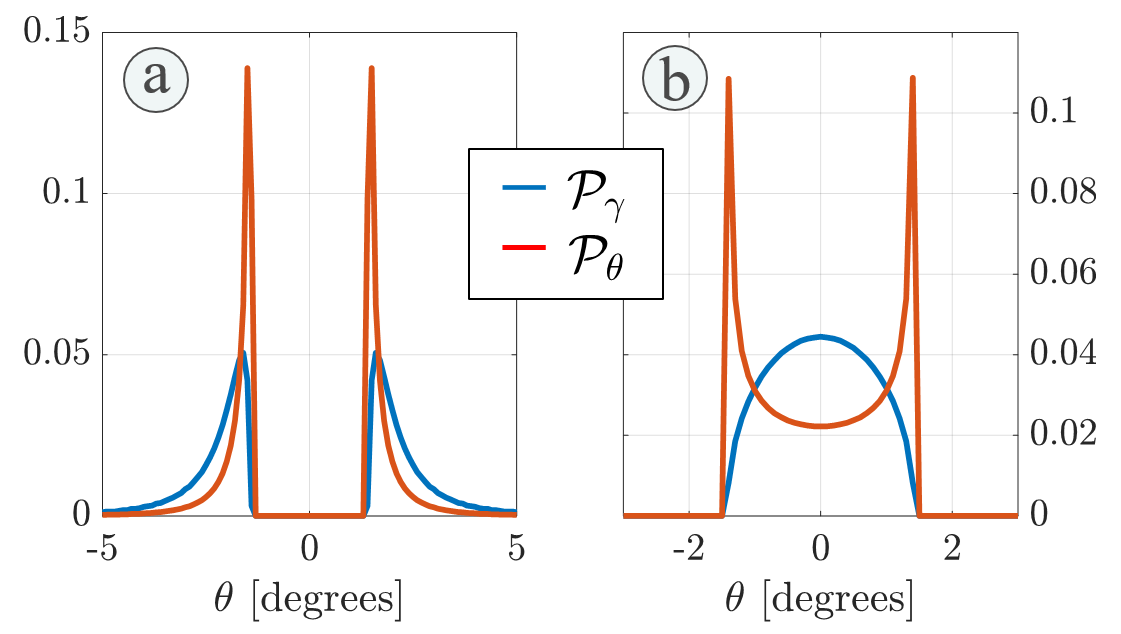}
    \caption{The angular probability distribution \({\cal P}_{\theta}\) and the angular distribution of emitted energy \({\cal P}_{\gamma}\) for two cases: (a) a free electron irradiated by only a plane electromagnetic wave in a vacuum, and (b) an electron moving along a magnetic field filament without laser fields. The corresponding trajectories are shown in \cref{fig:traj}. In panel (a), the normalized laser amplitude is \(a_0 = 80\), and the electron comes to a complete stop when the laser field reaches its maximum amplitude. In panel (b), the initial transverse and longitudinal components of the electron's momentum on the filament's axis are \(|p_{y,0}| = 80mc\) and \(p_{x,0} = 3200mc\), respectively, corresponding to an angle \(\theta_0 \approx 2.5 \times 10^{-2}\). The current density, which defines the magnetic filament, is set to \(j_0 = mc^3/\pi |e| \lambda_0^2\).}
    \label{fig: DLAfilamanet_dist_MATLAB}
\end{figure}

In the simulations discussed in \cref{sec: pic motivation}, the emitting electrons interact with both strong laser fields and the magnetic field of the filament as they move along it. To better understand how these interactions influence the emission process, we separately analyze the effects of the laser fields and the filament's magnetic field. Specifically, in this section, we examine the angular probability distribution \({\cal P}_{\theta}\) and the angular distribution of the emitted energy \({\cal P}_{\gamma}\) for two limiting cases: 1) an electron moving along the magnetic filament without any interaction with the laser fields, and 2) a free electron irradiated by only a plane electromagnetic wave in a vacuum. For simplicity, we assume that in both cases, the electron follows a flat trajectory, oscillating along the \(y\)-axis while moving forward along the \(x\)-axis. Examples of these trajectories are shown in \cref{fig:traj}. Our primary goal is to demonstrate that the plasma magnetic filament induces a single-lobe/peak emission profile. \Cref{fig: DLAfilamanet_dist_MATLAB} previews the \({\cal P}_{\theta}\) and \({\cal P}_{\gamma}\) results that we derive later in this section for the two cases under consideration.

In both cases, the problem reduces to examining the electron dynamics in prescribed electric and/or magnetic fields. The electron dynamics in given electric, $\bm{E}$, and magnetic, $\bm{B}$, fields is described by the following system of coupled differential equations:
\begin{align}
    &\frac{d \bm{p}}{d t} = - |e| \bm{E} - \frac{|e|}{\gamma m c} \left[ \bm{p} \times \bm{B} \right] + \bm{f}_{\rm{RF}}, \label{dpdt} \\
    &\frac{d \bm{r}}{d t} = \frac{c}{\gamma} \frac{\bm{p}}{m c}, \label{drdt} 
\end{align}
where $\bm{f}_{\rm{RF}}$ is the radiation friction force that accounts for the momentum loss due to photon emission. The force of radiation friction is anti-parallel to the electron momentum $\bm{p}$ in the case of an ultra-relativistic electron~\cite{landau.2013}, with
\begin{equation} 
    \bm{f}_{RF} = - \frac{8 \pi^2}{3} \frac{r_e}{\lambda_0} \frac{m c}{T} \rchi^2 \left( \frac{e B_{crit}}{m c \omega} \right)^2 \frac{\bm{p}}{p}, \label{EQ-RR2}
\end{equation}
where $r_e \equiv e^2 / m c^2 \approx 2.8 \times 10^{-13} \mbox{ cm}$ is the classical electron radius, $T \equiv 2 \pi / \omega$ is the laser period, and $\lambda_0 \equiv 2 \pi c / \omega$ is the laser wavelength in vacuum. The expression in \cref{EQ-RR2} is derived assuming that the energy of individual photons emitted by the electron is much smaller than the electron energy.

The emitted photon energy ${\cal E}_{\gamma}$ increases at the rate:
\begin{equation} \label{eq: emission 1}
    \frac{d {\cal E}_{\gamma}}{dt} = \left| \bm{f}_{\rm{RF}} \right| v.
\end{equation}
This can be derived by multiplying  equation~(\ref{dpdt}) by $\bm{p}$, resulting in the energy balance equation:
\begin{equation} \label{eq: energy balance}
    \frac{d \varepsilon_e}{d t} = - |e| (\bm{E} \cdot \bm{v}) - \left| \bm{f}_{\rm{RF}} \right| v,
\end{equation}
where $\varepsilon_e \hspace{-.5mm} = \hspace{-.5mm}(\gamma - 1)mc^2$ is the kinetic energy of the electron. The first term on the right-hand side represents the energy exchange with the prescribed electric field, indicating that the electron can gain or lose energy through its interactions with $\bm{E}$. The second term, which is always negative, corresponds to the energy loss due to photon emission. Since photons are emitted along $\bm{p}$, the expression given by \cref{eq: emission 1} can be used to determine the angular distribution of the emitted power. However, this requires knowledge of ${\cal P}_{\theta}$, which quantifies the amount of time the electron spends oriented at a specific angle $\theta$.

To determine \({\cal P}_{\theta}\) over a full transverse oscillation, we take advantage of the symmetry in the electron's motion. The trajectories are shown in \cref{fig:traj} to clarify the geometric considerations used in the derivations. For both cases, we divide the complete transverse oscillation into four equal quarters, as illustrated in \cref{fig:traj}a and \cref{fig:traj}b. This division creates four segments where \(\theta(t)\) is a single-valued function, meaning \(\theta\) either consistently increases or decreases within each segment. We then introduce the probability distribution for each segment, \({\cal P}_{\theta}^{(i)}\), where \(i\) represents the segment number. The distribution \({\cal P}_{\theta}^{(i)}\) is normalized according to
\begin{equation} \label{eq: normalization}
    \int {\cal P}_{\theta}^{(i)} (\theta) d\theta = 1,
\end{equation}
with the integration performed over the corresponding \(\theta\) range for that specific segment. 

We now focus on a single segment where \(\theta > 0\) and \(d\theta / dt \geq 0\), denoting relevant quantities with a superscript \(+\). This corresponds to the first segment in \cref{fig:traj}a and the second segment in \cref{fig:traj}b, hence the use of the superscript \(+\) instead of a number. The angular probability distribution for this quarter is expressed as:
\begin{equation} \label{def: p theta v1}
    {\cal P}_{\theta}^{+} (\theta) = \frac{4}{\tau } \left( \frac{d \theta}{dt} \right)^{-1},
\end{equation}
where \(\tau\) is the period of the full oscillation. This expression satisfies the normalization condition given by \cref{eq: normalization}. This can be verified by converting the integration to \(t\), noting that \(\theta\) is a single-valued function, and then using the fact that the integration interval over \(t\) for a single quarter is \(\tau/4\). To obtain the actual form of \({\cal P}_{\theta}^{+} (\theta)\), the function \(d\theta / dt\) must be calculated from the solution of the equations of motion. Due to the symmetry of the oscillations, the angular probability distribution for the quarter where \(\theta > 0\) and \(d\theta / dt \leq 0\) is identical to \({\cal P}_{\theta}^{+} (\theta)\). The electron spends half of its time with \(\theta > 0\), which leads us to conclude that ${\cal P}_{\theta} = {\cal P}_{\theta}^{+} (\theta) / 2$ for \(\theta > 0\). To determine the angular probability distribution for the entire electron trajectory, we take into account that ${\cal P}_{\theta} (-\theta) = {\cal P}_{\theta} (\theta)$, resulting in:
\begin{equation} \label{def: p theta}
    {\cal P}_{\theta} (\theta) = \begin{cases}
{\cal P}_{\theta}^{+} (\theta) / 2&\text{for $\theta \geq 0$,}\\
{\cal P}_{\theta}^{+} (-\theta) / 2 &\text{for $\theta < 0$.}
\end{cases}
\end{equation}

To determine ${\cal P}_{\gamma}$, we use a similar approach by first considering a single quarter of the electron oscillation where $\theta > 0$ and $d \theta / dt \geq 0$. In this quarter, $\theta(t)$ is a single-valued function, allowing us to rewrite \cref{eq: emission 1} as:
\begin{equation}
        \frac{d {\cal E}_{\gamma}}{d \theta} \frac{d \theta}{dt} = \left| \bm{f}_{\rm{RF}} \right| v.
\end{equation}
Utilizing the definition of ${\cal P}_{\theta}^{+} (\theta)$ from \cref{def: p theta v1}, we obtain
\begin{equation} \label{def: emitted energy}
    \frac{d {\cal E}_{\gamma}}{d \theta} = \frac{\tau }{4} {\cal P}_{\theta}^{+} (\theta) \left| \bm{f}_{\rm{RF}} \right| v.
\end{equation}
For this quarter, the angular distribution of the emitted energy is:
\begin{equation} \label{Q+}
    {\cal P}_{\gamma}^{+} (\theta) = \frac{4}{{\cal E}_{\gamma}^{tot}} \frac{d {\cal E}_{\gamma}}{d \theta} = \frac{\tau}{{\cal E}_{\gamma}^{tot}}  {\cal P}_{\theta}^{+} (\theta) \left| \bm{f}_{\rm{RF}} \right| v,
\end{equation}
where ${\cal E}_{\gamma}^{tot}$ is the total energy emitted over the entire oscillation. The distribution is normalized such that 
\begin{equation}
    \int {\cal P}_{\gamma}^{+} (\theta) d \theta = 1,
\end{equation}
with the integration performed over the $\theta$ range corresponding to the considered segment. To find the angular distribution of the energy emitted over the entire electron trajectory, we again consider the symmetry of the oscillations. The angular distribution of the emitted energy for the quarter where $\theta > 0$ and $d \theta / dt \leq 0$ is identical to ${\cal P}_{\gamma}^{+} (\theta)$. Since the electron spends half of its time with $\theta > 0$, we have:
\begin{equation} \label{def: q theta}
    {\cal P}_{\gamma} (\theta) = \begin{cases}
{\cal P}_{\gamma}^{+} (\theta) / 2&\text{for $\theta \geq 0$,}\\
{\cal P}_{\gamma}^{+} (-\theta) / 2 &\text{for $\theta < 0$.}
\end{cases}
\end{equation}

\Cref{Q+} for ${\cal P}_{\gamma}^{+} (\theta)$ offers key insights into the angular dependence of photon emission. To highlight this, we first replace $v$ with $c$. The angular dependence on the right-hand side is then determined solely by the product of ${\cal P}_{\theta}^+$ and $\left| \bm{f}_{\rm{RF}} \right|$. The amplitude of the radiation friction force inherits its angular dependence from $\rchi^2$. We thus conclude that
\begin{equation} \label{eq: simple Q_gamma}
    {\cal P}_{\gamma}^+ \propto {\cal P}_{\theta}^+ \rchi^2.
\end{equation}
This shows that ${\cal P}_{\gamma}$, defined using ${\cal P}_{\gamma}^+$, can have a single lobe even if ${\cal P}_{\theta}$ has two lobes. However, this requires a single-peaked  $\rchi^2$ that can counteract the two-lobed shape of ${\cal P}_{\theta}$. In general, ${\cal P}_{\theta}$ is influenced by $\bm{f}_{RF}$, complicating the analysis of angular dependence. Typically, the role of the radiation friction force increases with the laser $a_0$. \mc{In the regimes considered in \cref{sec: pic motivation}, the value of $a_0$ is relatively modest, resulting in weak feedback from the radiation friction force. To simplify the analysis in this section, we neglect this effect and adopt an approach where ${\cal P}_{\theta}$ and $\rchi$ are calculated without including $\bm{f}_{RF}$ in Eq.~(\ref{dpdt}).}

Next, we examine the two scenarios previously mentioned, starting with the case of an electron moving along a magnetic filament with an azimuthal magnetic field, as shown in \cref{fig:traj}a. We consider an electron following a flat trajectory in the \((x,y)\) plane that intersects the axis of the magnetic filament at \(z = 0\). The magnetic field $B_z$ experienced by the electron varies with $y$, changing sign at $y = 0$, such that $B_z < 0$ for $y > 0$ and $B_z > 0$ for $ y < 0$. 

Along the considered trajectory, the electron's momentum can be expressed as:
\begin{equation}
    \bm{p} = \bm{e}_x p \cos \theta + \bm{e}_y p \sin \theta,
\end{equation}
where \(p\) is the magnitude of the momentum and \(\theta\) is the angle between the momentum vector and the axis of the filament. Since the electron's energy is conserved, \(p\) remains constant during its motion. Given that the electron experiences a magnetic field directed along the \(z\)-axis, we derive from Eq.~(\ref{dpdt}) that
\begin{equation} \label{dtheta-filament}
    \frac{d \theta}{d t} = \frac{|e| B_z}{\gamma m c}.
\end{equation}
We now consider an electron crossing the axis with \(\theta = \theta_0\), where \(0 < \theta_0 < \pi/2\). At \(y > 0\), the filament's magnetic field \(B_z\) is negative, causing the electron to be deflected in the positive \(x\)-direction. According to \cref{dtheta-filament}, the angle decreases from \(\theta_0\) to \(-\theta_0\) while the electron moves above the axis. After returning to \(y = 0\), the electron crosses the axis and enters the region where \(B_z > 0\). At this point, the angle \(\theta\) begins to increase, rising from \(-\theta_0\) back to \(\theta_0\). Therefore, the electron undergoes transverse oscillations as it progresses along the filament.

We are now in a good position to determine how ${\cal P}_{\theta}^{+}$ and ${\cal P}_{\gamma}^{+}$ scale with $|B_z|$. In the quarter where $\theta > 0$ and $d \theta / dt \geq 0$, we have $B_z \geq 0$. According to \cref{def: p theta v1}, we find:
\begin{equation} \label{P for filament}
    {\cal P}_{\theta}^{+} = \frac{4}{\tau } \left( \frac{d \theta}{dt} \right)^{-1} = \frac{4}{\tau } \frac{\gamma m c}{|e| B_z} \propto 1/B_z.
\end{equation}
From \cref{chi}, we also readily find:
\begin{equation} \label{chi-filament}
    \rchi = \frac{p}{m c} \frac{|B_z|}{B_{crit}}.
\end{equation}
It then follows from \cref{eq: simple Q_gamma} that:
\begin{equation}
    {\cal P}_{\gamma}^{+} \propto {\cal P}_{\theta}^{+} \rchi^2 \propto |B_z|.
\end{equation}
Finally, using the symmetry of the trajectory, we arrive at the key scaling:
\begin{equation} \label{Q for filament}
    {\cal P}_{\gamma} \propto |B_z|.
\end{equation}

To determine the actual dependence of ${\cal P}_{\theta}^{+}$ and ${\cal P}_{\gamma}^{+}$ on $\theta$, we must first find \( y(\theta) \), which requires finding the electron trajectory. We consider a filament with a magnetic field given by:
\begin{equation} \label{B-filament}
    B_z = - 2 \pi y j_0/c.
\end{equation}
This field is generated by a current with uniform current density \( j_x = -j_0 \), where \( j_0 > 0 \). For simplicity, we assume \( |\theta| \ll 1 \) and that the electron is ultra-relativistic, reducing the transverse component of Eq.~(\ref{drdt}) to:
\begin{equation} \label{eq: dydt}
    \frac{dy}{dt} \approx c \theta. 
\end{equation}
Taking the time derivative of \cref{dtheta-filament} and substituting the expression for \( B_z \) from \cref{B-filament} and \( dy/dt \) from \cref{eq: dydt}, we obtain:
\begin{equation} 
    \frac{d^2 \theta}{d t^2} + \Omega^2 \theta = 0,
\end{equation}
where:
\begin{equation} \label{Omega}
    \Omega = \sqrt{\frac{2 \pi}{\gamma} \frac{|e| j_0}{mc}}.
\end{equation}
The solution to this equation, for an electron crossing the axis at \( t = 0 \) while moving in the positive \( y \)-axis direction, is:
\begin{equation} \label{eq: 30}
    \theta = \theta_0 \cos (\Omega t).
\end{equation}
From \cref{eq: dydt}, we find:
\begin{equation} \label{eq: 31}
    y = \frac{c \theta_0}{\Omega} \sin (\Omega t).
\end{equation}
The final step is to express \( |y| \) in terms of \( \theta \). Using \cref{eq: 31} and \cref{eq: 30}, we find:
\begin{equation} \label{eq: 32}
    |y(\theta)| = \frac{c \theta_0}{\Omega}  |\sin (\Omega t)| = \frac{c \theta_0}{\Omega}  \sqrt{1 - \cos^2 (\Omega t)} = \frac{c}{\Omega} \sqrt{\theta_0^2 - \theta^2}.
\end{equation}

We now have all the elements needed to determine \( {\cal P}_{\theta} (\theta) \) and \( {\cal P}_{\gamma} (\theta) \). From \cref{B-filament} and \cref{eq: 32}, we obtain:
\begin{equation} 
    |B_z| \propto |y| \propto \sqrt{\theta_0^2 - \theta^2 }.
\end{equation}
Using this relation in \cref{P for filament} and substituting the result into \cref{def: p theta}, we find:
\begin{equation}
    {\cal P}_{\theta} (\theta) \propto \left[ \theta_0^2 - \theta^2 \right]^{-1/2}.
\end{equation}
The resulting angular probability distribution has two peaks at \( \theta = \theta_0 \) and \( \theta = -\theta_0 \), corresponding to the electron crossing the axis of the magnetic filament. Using \cref{Q for filament}, we find:
\begin{equation} 
    {\cal P}_{\gamma} (\theta) \propto \left[ \theta_0^2 - \theta^2 \right]^{1/2}.
\end{equation}
This angular distribution of the emitted power has a single peak at \( \theta = 0 \). According to \cref{eq: 30} and \cref{eq: 31}, \( \theta = 0 \) occurs at the transverse turning points located at \( y = \pm c \theta_0/\Omega \).

\Cref{fig: DLAfilamanet_dist_MATLAB} shows an exact results for $\cal{P}_{\theta}$ and $\cal{P}_{\gamma}$ obtained by solving Eqs.~(\ref{dpdt}) and (\ref{drdt}) and determining the trajectory shown in \cref{fig:traj}a. In this example, the transverse and longitudinal components of the electron momentum on the filament's axis are \( |p_y| = 80 mc \) and \( p_x = 3200 mc \), corresponding to \( \theta_0 \approx 2.5 \times 10^{-2} \approx 1.4^{\circ}\). The current density is set to \( j_0 = mc^3/ \pi |e| \lambda_0^2 \). Although there is no laser in this problem, we used \( \lambda_0 \) to define \( j_0 \) to facilitate comparison with setups where the filament is driven by a laser. As shown in \Cref{fig: DLAfilamanet_dist_MATLAB}, the emission from an electron moving along the filament without a laser present has only a single lobe at \( \theta = 0 \), even though the angular probability peaks at \( \theta = \pm \theta_0 \).

To conclude this section, we turn our attention to another straightforward but informative example: a free electron irradiated by a plane electromagnetic wave in a vacuum. The general solution for this scenario is well-known~\cite{Acharya.IEEETPS.1993.VLA}, so we will not go through the derivation, but we will present the key results necessary for evaluating the angular dependence of ${\cal P}_{\theta}$ and ${\cal P}_{\gamma}$.

We consider a wave with a normalized vector potential given by
\begin{equation}
    \bm{a} = \bm{e}_y a = \bm{e}_y a_0 \sin(\xi),
\end{equation}
where \( a_0 \) is the normalized wave amplitude and
\begin{equation} \label{xi}
    \xi = \omega (t - x/c)
\end{equation}
is a dimensionless phase variable. The electric field of the wave has only a \( y \)-component:
\begin{equation}
    \frac{|e| E_y}{m \omega c} = -  \frac{da}{d\xi} = -a_0 \cos(\xi).
\end{equation}
The magnetic field is directed along the \( z \)-axis, with \( B_z = E_y \). In this wave, the electron possesses two integrals of motion:
\begin{align}
    p_y/mc - a &= C_1, \\
    \gamma - p_x / mc &= C_2.
\end{align}
We consider an electron initially at rest (\( p_x = p_y = 0 \)) at \( \xi = 0 \). Thus, we have \( C_1 = 0 \) and \( C_2 = 1 \). Using these integrals of motion, we find:
\begin{align}
    p_y/mc &= a, \label{sol-1} \\
    p_x/mc &= a^2/2. \label{sol-2} 
\end{align}
This solution indicates that the electron moves along a sideways parabola in momentum space. The polar angle defining the direction of \( \bm{p} \) (the polar angle in momentum space) along this parabola is given by:
\begin{equation} \label{theta-DLA}
    \theta = \tan^{-1} \left( 2/a \right).
\end{equation}

The derived expression for \(\theta\) allows us to determine the key features of the angular dependence of \({\cal P}_{\theta}\) and \({\cal P}_{\gamma}\). We first consider the part of the trajectory where \(a \geq 0\) and \(a\) decreases from its maximum value of \(a_0\) to 0 (second segment in \cref{fig:traj}b). At \(a = a_0\), the electron has the smallest angle:
\begin{equation}
    \theta = \theta_{\min} \equiv \tan^{-1} \left( 2/a_0 \right).
\end{equation}
As \(a\) decreases, the angle increases and reaches \(\pi/2\). The key observation here is that \(\theta \geq \theta_{\min}\). Similarly, for the part of the trajectory where \(a < 0\) (third and fourth segments in \cref{fig:traj}b), we find that \(\theta \leq -\theta_{\min}\). Because \(|\theta| \geq \theta_{\min}\) and the trajectory is symmetrical, it is impossible for \({\cal P}_{\theta}\) and \({\cal P}_{\gamma}\) to have just a single peak. Instead, they must both have two peaks: one at a negative \(\theta\) and another at a positive \(\theta\). \Cref{fig: DLAfilamanet_dist_MATLAB} confirms this by presenting an exact solution obtained by solving Eqs.~(\ref{dpdt}) and (\ref{drdt}) for \(a_0 = 80\).

To conclude this section, we  compare the shapes of \({\cal P}_{\theta} (\theta)\) and \({\cal P}_{\gamma} (\theta)\) in \cref{fig: DLAfilamanet_dist_MATLAB} with those obtained from our PIC simulations, as shown in \cref{fig: dist pic}. When the electron dynamics is governed by the laser, both \({\cal P}_{\theta} (\theta)\) and \({\cal P}_{\gamma} (\theta)\) exhibit two peaks, similar to the patterns observed in the simulation with \(n_{e,0} = 4 n_{cr}\). In contrast, when electron dynamics is influenced by the magnetic field of the filament, \({\cal P}_{\theta} (\theta)\) shows two peaks, while \({\cal P}_{\gamma} (\theta)\) displays a single peak. This behavior is akin to what we see in the simulation with \(n_{e,0} = 1n_{cr}\). We therefore conjecture that the magnetic filament plays a crucial role in establishing the single-lobe emission profile observed in the simulation with \(n_{e,0} = 1n_{cr}\).

%*****************************

\section{Emission of electrons undergoing direct laser acceleration in a test-electron model} \label{sec: test electron model}

In \cref{sec: simple estimates}, we explored photon emission in two limiting cases: one where the electron is influenced solely by the laser fields, and another where it is influenced solely by the plasma magnetic field. However, in the simulations presented in \cref{sec: pic motivation}, the motion of the emitting electrons is a combination of oscillations induced by the magnetic filament and those induced by the laser. To draw a closer parallel with these simulations, we now analyze the electron dynamics in the presence of both laser and plasma fields using a test-electron model~\cite{Khudik.PoP.2016.Universal, jirka.njp.2020, Yeh.NJP.2021.Friction, Arefiev.PoP.2024.Modulation} that treats these fields as prescribed. \mc{This approach, which uses prescribed rather than self-consistently calculated fields, allows us to clearly identify the features of the electron dynamics responsible for changes in the emission pattern.}
%Treating the fields as prescribed rather than calculating them self-consistently allows us to reliably identify the features of the electron dynamics responsible for the changes in the emission pattern.

In our test-electron model, the electron dynamics is governed by Eqs.~(\ref{dpdt}) and (\ref{drdt}), \rc{which include the radiation friction force associated with photon emission.} The electric and magnetic fields are represented as a superposition of oscillating laser fields (denoted by the superscript ``laser") and static plasma fields (denoted by the superscript ``pl"): $\bm{E} = \bm{E}^{\rm{laser}} + \bm{E}^{\rm{pl}}$ and $\bm{B} = \bm{B}^{\rm{laser}} + \bm{B}^{\rm{pl}}$. The selection of these fields is informed by the PIC simulations discussed in \cref{sec: pic motivation}. The following three paragraphs detail how we approximate the observed fields within our model. For simplicity, we focus on electrons with flat trajectories confined to the \((x,y)\) plane that intersects the axis of the magnetic filament at \(z = 0\), and therefore limit our analysis to the field profiles within this plane.

We approximate the plasma magnetic field observed in the simulations as a static field generated by a current filament with a uniform current density \(j_x = -j_0\) (where \(j_0 > 0\)). The magnetic field in the \((x,y)\) plane is given by
\begin{equation}
    B_z^{\rm{pl}} = - \frac{m_e c^2}{|e|} \frac{2 \alpha y}{\lambda_0^2},
\end{equation}
where we introduce the dimensionless parameter
\begin{equation} \label{eq: def alpha}
    \alpha \equiv \pi \lambda_0^2 j_0/J_A,
\end{equation}
which represents the ratio of the current flowing through a circular area with radius \(\lambda_0\) to the classical Alfvén current \(J_A = m_e c^3/|e|\). Note that this expression for \(B_z^{\rm{pl}}\) is consistent with \cref{B-filament} in \cref{sec: simple estimates}. As discussed in that section, the current density \(j_0\) determines the frequency \(\Omega\) of the transverse electron oscillations induced by the plasma magnetic field. The expression for \(\Omega\) from \cref{Omega} can be rewritten in terms of \(\alpha\) as
\begin{equation}
    \Omega / \omega = \sqrt{\alpha / 2 \pi^2 \gamma}.
\end{equation}

For simplicity, our model neglects plasma electric fields. In the regime of interest, the electric field is primarily radial and can induce transverse electron oscillations. However, our simulations show that the force from \(\bm{B}^{\rm{pl}}\) is significantly more substantial than that from \(\bm{E}^{\rm{pl}}\). This is commonly observed in dense plasmas (\(n_e \gtrsim n_{cr}\)) irradiated by ultra-high-intensity lasers (\(a_0 \gg 1\))~\cite{Jansen.PPCF.2018.Pair}. Consequently, we disregard \(\bm{E}^{\rm{pl}}\) in our analysis. While it can be included if needed, this would only affect the frequency of the transverse oscillations induced by the plasma without qualitatively altering the electron dynamics.

In the PIC simulations, the laser beam is guided by the plasma through a channel created by the laser itself. This channel prevents the beam from diverging, allowing it to propagate through the plasma over distances greater than its diffraction length without significant amplitude loss. The plasma within the channel also causes the laser phase velocity, $v_{ph}$, to become slightly superluminal. Although the difference between \(v_{ph}\) and \(c\) is usually small in the regime of interest, 
\begin{equation}
    \delta u \equiv \frac{v_{ph} - c}{c} \ll 1,
\end{equation}
this small deviation plays a crucial role in determining electron energy gain during direct laser acceleration~\cite{Khudik.PoP.2016.Universal}, \mc{as it dictates how long the electron can continue to gain energy from the laser. This is why we retain this parameter and later use it as a scan parameter in our analysis.}

Our model is designed to capture two key aspects of laser propagation observed in the PIC simulations: a nearly constant laser amplitude over a long propagation distance and a superluminal phase velocity. To achieve this, we approximate the laser fields as a linearly polarized plane wave with a superluminal phase velocity \( v_{ph} \). The expressions for the fields are given by:
\begin{align}
    E_y^{\rm{laser}} &= E_0 \cos(\xi), \label{E-laser} \\
    B_z^{\rm{laser}} &= E_y/u, \label{B-laser}
\end{align}
where \( E_0 \) is the laser amplitude,
\begin{equation}
    u \equiv v_{ph}/c
\end{equation}
is the normalized phase velocity, and
\begin{equation} \label{xi-u}
    \xi = \omega (t - x u/c)
\end{equation}
is a dimensionless phase variable. 

\mc{In the considered setup, an electron can enter a regime where it steadily gains energy from the laser, despite the oscillations of the laser field at its location. This prolonged energy gain requires frequency matching between the transverse oscillations induced by the plasma magnetic field (\(\Omega\)) — commonly referred to as betatron oscillations — and the oscillations of the laser electric field at the electron’s location (\(\omega'\)). As the electron gains energy, its forward motion becomes ultra-relativistic, significantly reducing \(\omega'\). Typically, \(\omega'\) is much smaller than the laser frequency \(\omega\), which enables the required frequency matching with \(\Omega\).  }

\mc{Sustaining the frequency matching is complicated by the fact that both \(\Omega\) and \(\omega'\) depend on the electron's energy, but they scale differently with \(\gamma\)~\cite{Khudik.PoP.2016.Universal, Arefiev.PoP.2024.Modulation}. This difference in scaling causes frequency detuning as the electron gains energy. Counterintuitively, this detuning can be mitigated by introducing an appropriate degree of superluminosity, represented by \(\delta u\)~\cite{Khudik.PoP.2016.Universal}. As a result, the energy gain can be further prolonged, allowing the electron to achieve significantly higher energies than in cases where the difference between \(v_{ph}\) and \(c\) is neglected.  }

\rc{It is therefore natural to distinguish two qualitatively different regimes, which we refer to as \emph{efficient} DLA and \emph{inefficient} DLA. In inefficient DLA, the electron never achieves frequency matching, leading to strong modulations of its \(\gamma\)-factor, a key identifier of this regime. In efficient DLA, the electron achieves frequency matching, enabling sustained energy gain over multiple laser oscillations. A signature of this significant energy gain is the reduced modulation of \(\gamma\). In this study, we focus specifically on photon emission in these two regimes. As the requirements for frequency matching have been extensively studied in the literature~\cite{Khudik.PoP.2016.Universal, Arefiev.PoP.2016.Beyond, Arefiev.JPP.2015.Novel, Babjak.NJP.2024.Density, Babjak.PRL.2024.GeV}, we omit the derivations and concentrate on the emission characteristics.}

\begin{figure}
    \centering
    \includegraphics[width=.75\linewidth]{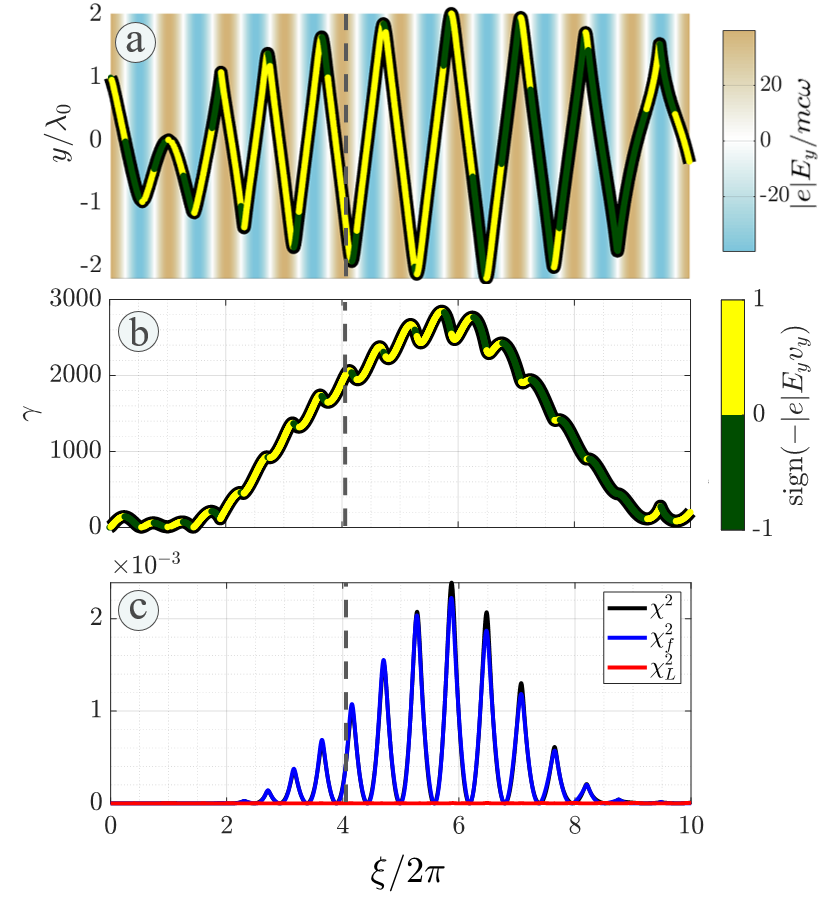}
    \caption{Regime with \emph{efficient} DLA: (a) Electron trajectory plotted over the transverse laser electric field; (b) relativistic \(\gamma\)-factor; (c) parameter \(\rchi\) computed using three different sets of fields. The black curve, \(\rchi^2\), is computed using \(\bm{E} = \bm{E}^{\rm{laser}} + \bm{E}^{\rm{pl}}\) and \(\bm{B} = \bm{B}^{\rm{laser}} + \bm{B}^{\rm{pl}}\); the blue curve, \(\rchi_f^2\), is computed using \(\bm{E} = 0\) and \(\bm{B} = \bm{B}^{\rm{pl}}\); the red curve (\(\rchi_L^2\)) is computed using \(\bm{E} = \bm{E}^{\rm{laser}}\) and \(\bm{B} = \bm{B}^{\rm{laser}}\). The color coding along the trajectory in (a) and (b) indicates the sign of \(E_y v_y\), with yellow representing energy gain and dark green representing energy loss. \rc{The impact of the radiation friction becomes significant past the vertical dashed line.}  Refer to the `Efficient DLA' column in Table~\ref{table: efficient and inefficient} for the complete set of parameters used to generate this trajectory.}
    \label{fig: efficient_particle}
\end{figure}

\Cref{fig: efficient_particle} presents an example of a regime where DLA is particularly efficient due to prolonged frequency matching facilitated by the superluminosity. All parameters necessary to reproduce our result are provided in \cref{table: efficient and inefficient} (see the `Efficient DLA' column). \Cref{fig: efficient_particle}a shows the electron trajectory plotted over the laser electric field, using \(\xi\) (rather than \(x\)) as the variable along the horizontal axis. This choice facilitates a comparison between the electron's oscillations and those of the laser electric field. The color-coding along the electron trajectory represents the sign of \(-|e| E_y v_y\), with yellow indicating energy gain and dark green indicating energy loss. In the absence of the plasma magnetic field, these colors would alternate. A clear signature of the efficient DLA induced by the plasma magnetic field is that energy gain (yellow) dominates over multiple  transverse oscillations (approximately five in this case). The corresponding \(\gamma\) profile is shown in \cref{fig: efficient_particle}b, where it reaches a peak value of almost 3000 — two orders of magnitude higher than \(a_0\).

\begin{table}
\footnotesize
\centering
\begin{tabular}{ | p{6cm} | p{3.5cm}| p{3.5cm} |}
\hline
\multicolumn{3}{|c|}{Test-electron model parameters} \\
\hline
& Inefficient DLA & Efficient DLA \\
\hline
Initial position $(t = 0)$ & $x = 0~\micron$, $y = 1~\micron$ & $x = 0~\micron$, $y = 1~\micron$ \\
\hline
Initial momentum & $p/mc = 0$  & $p/mc = 0$  \\
\hline
Normalized laser amplitude, $a_{0}$ & 40 & 40\\
\hline
E-field peak amplitude, $E_{0}$ & 1.29 $\times 10^{14}$~V/m& 1.29 $\times 10^{14}$~V/m \\
\hline
Wavelength, $\lambda_{0}$ & $1.00~\micron$ & $1.00~\micron$ \\
\hline
Laser frequency & $\omega = 2 \pi c / \lambda_0$ & $\omega = 2 \pi c / \lambda_0$ \\
\hline
Normalized current density, $\alpha$  & 1.0 & 12.0 \\
\hline
Normalized phase-velocity, $v_{ph}/c$ & 1.01 & 1.01 \\
\hline
Propagation & +$\bm{x}$ direction & +$\bm{x}$ direction \\
\hline
Polarization & linear, with $E_y$ and $B_z$ & linear, with $E_y$ and $B_z$\\
\hline
Temporal profile & uniform & uniform \\
\hline
\end{tabular}
\caption{Parameters used in the test-electron model of \cref{sec: test electron model} to generate \cref{fig: inefficient_particle} (inefficient DLA regime) and \cref{fig: efficient_particle} (efficient DLA regime).}
\label{table: efficient and inefficient}
\end{table}

\begin{figure}[b!]
    \centering
    \includegraphics[width=0.7\linewidth]{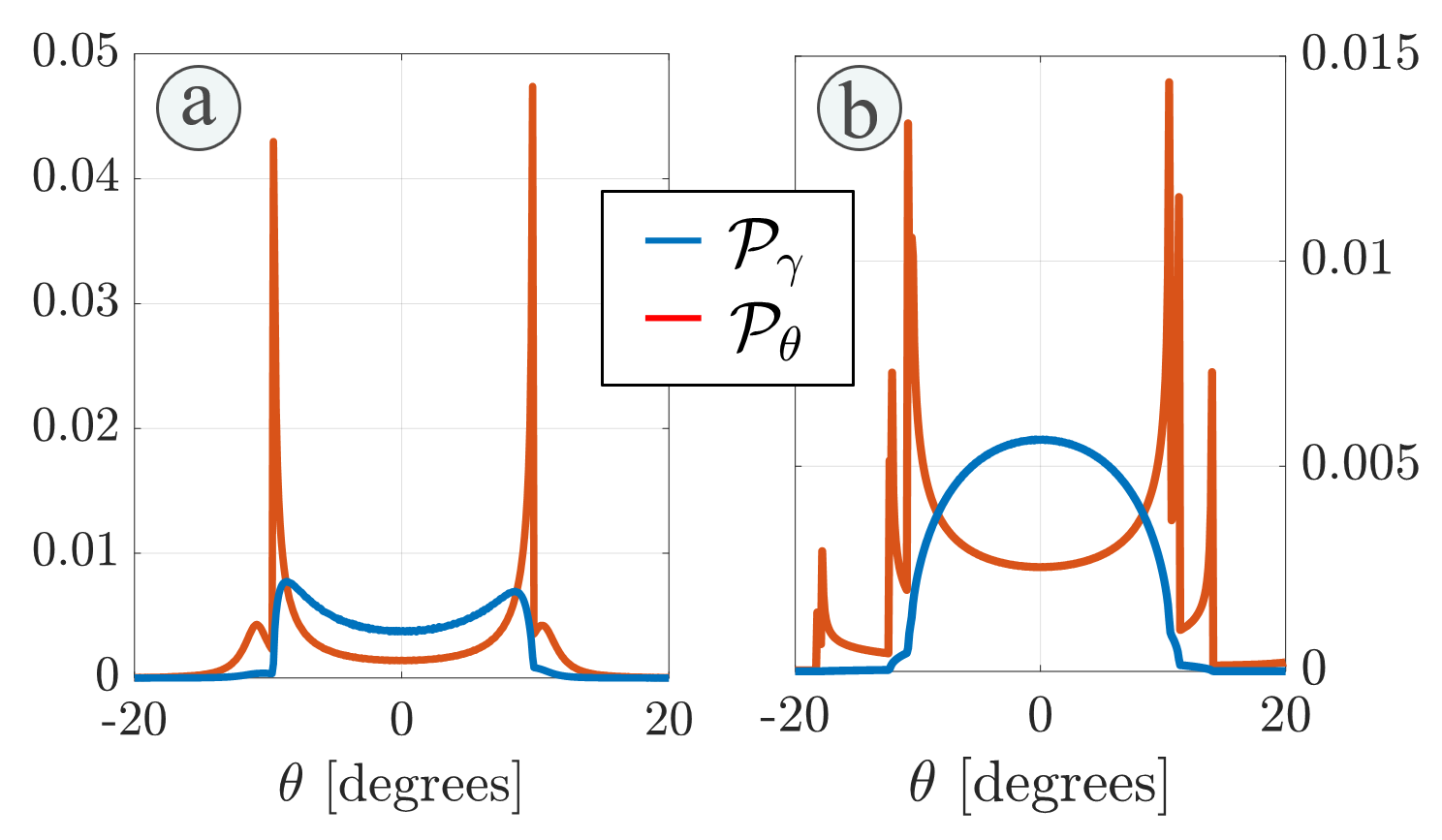}
    \caption{The angular probability distribution \({\cal P}_{\theta}\) and the angular distribution of emitted energy \({\cal P}_{\gamma}\) for two regimes: (a) regime of \emph{inefficient} DLA from \cref{fig: inefficient_particle} and (b) regime of \emph{efficient} DLA from \cref{fig: efficient_particle}.}
    \label{fig: dist matlab}
\end{figure}

\Cref{fig: dist matlab}b displays both the angular probability distribution \({\cal P}_{\theta}\) and the angular distribution of the emitted energy \({\cal P}_{\gamma}\) for the trajectory shown in \cref{fig: efficient_particle}a. As anticipated, \({\cal P}_{\theta}\) exhibits two primary lobes or peaks, with smaller peripheral spikes caused by variations in the angle at which the electron crosses the axis of the filament as it gains or loses energy. In contrast, \({\cal P}_{\gamma} (\theta)\) shows a single peak at \(\theta = 0\), closely resembling \({\cal P}_{\gamma} (\theta)\) in \cref{fig: DLAfilamanet_dist_MATLAB}b, which corresponds to an electron emitting photons while moving along the filament without the influence of the laser.

The similarities between the efficient DLA regime and the limiting case of an electron moving along the filament, as discussed in \cref{sec: simple estimates}, extend beyond the \({\cal P}_{\gamma} (\theta)\) distribution. During efficient DLA, the overall energy gain across multiple oscillations is substantial, so the relative change in \(\gamma\) during each betatron oscillation becomes small compared to the total value of \(\gamma\). This is analogous to the limiting case in \cref{sec: simple estimates}, where the \(\gamma\)-factor was treated as constant.

We assess the impact of the plasma magnetic field on electron emission by computing \(\rchi\) in three different ways. The black curve in \cref{fig: efficient_particle}c shows \(\rchi\) calculated using \(\bm{E} = \bm{E}^{\rm{laser}} + \bm{E}^{\rm{pl}}\) and \(\bm{B} = \bm{B}^{\rm{laser}} + \bm{B}^{\rm{pl}}\). The blue curve represents \(\rchi_f\), the value of \(\rchi\) computed by considering only the magnetic field of the filament, i.e., \(\bm{E} = 0\) and \(\bm{B} = \bm{B}^{\rm{pl}}\). Finally, the red curve depicts \(\rchi_L\), the value of \(\rchi\) calculated using only the laser field, i.e., \(\bm{E} = \bm{E}^{\rm{laser}}\) and \(\bm{B} = \bm{B}^{\rm{laser}}\). The fact that \(\rchi_f\) and \(\rchi\) are almost indistinguishable confirms that the emission process during efficient DLA is predominantly governed by the magnetic field of the filament. The role of the laser field is negligible, as evident from the plot of \(\rchi_L\). We can thus conclude that the emission in the efficient DLA regime closely resembles the limiting case of an electron moving along the filament, as discussed in \cref{sec: simple estimates}.

\begin{figure}
    \centering
    \includegraphics[width=.75\linewidth]{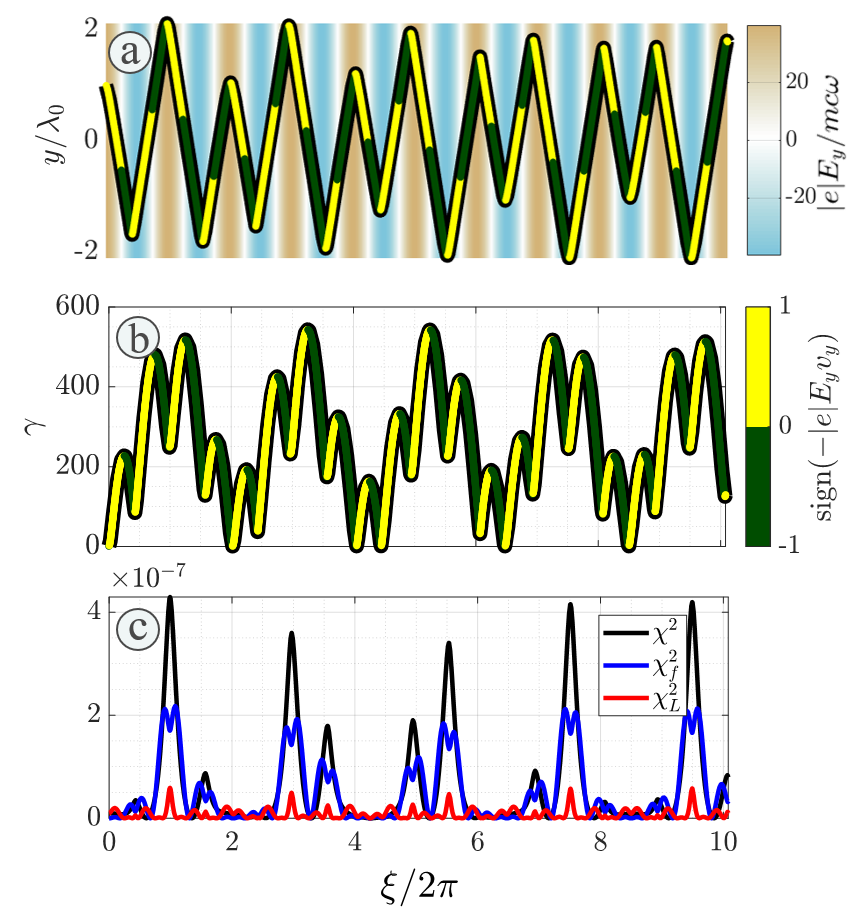}
    \caption{Regime with \emph{inefficient} DLA: (a) Electron trajectory plotted over the transverse laser electric field; (b) relativistic \(\gamma\)-factor; (c) parameter \(\rchi\) computed using three different sets of fields. The black curve, \(\rchi^2\), is computed using \(\bm{E} = \bm{E}^{\rm{laser}} + \bm{E}^{\rm{pl}}\) and \(\bm{B} = \bm{B}^{\rm{laser}} + \bm{B}^{\rm{pl}}\); the blue curve, \(\rchi_f^2\), is computed using \(\bm{E} = 0\) and \(\bm{B} = \bm{B}^{\rm{pl}}\); the red curve (\(\rchi_L^2\)) is computed using \(\bm{E} = \bm{E}^{\rm{laser}}\) and \(\bm{B} = \bm{B}^{\rm{laser}}\). The color coding along the trajectory in (a) and (b) indicates the sign of \(E_y v_y\), with yellow representing energy gain and dark green representing energy loss. Refer to the `Inefficient DLA' column in Table~\ref{table: efficient and inefficient} for the complete set of parameters used to generate this trajectory.}
    \label{fig: inefficient_particle}
\end{figure}

To determine under what conditions the laser's influence on emission is negligible, we examine the expression for \(\chi\) given by \cref{chi}, which can be recast as \(\chi = \gamma {\cal{F}} /  B_{\text{crit}}\), where 
\begin{equation} \label{E-effective}
    {\cal{F}} \equiv \left| \bi{E} - \frac{\bi{p} (\bi{p} \cdot \bi{E})}{p^2} + \frac{1}{\gamma m c} \left[ \bi{p} \times \bi{B} \right] \right|.
\end{equation}
Since the \(\gamma\)-factor changes insignificantly during each oscillation, peaks in \(\chi\) directly correspond to peaks in \({\cal{F}}\). A comparison of \cref{fig: efficient_particle}c with \cref{fig: efficient_particle}a clearly shows that these peaks occur at turning points (\(p_y = 0\)).  

Next, we consider a single turning point, denoting the plasma magnetic field at this point as $B_*$. By definition of a turning point, the $(\bi{p} \cdot \bi{E})$-term vanishes. When the laser field reaches its maximum amplitude, we have
\begin{equation} \label{E-effective-2}
    {\cal{F}} \approx \left| E_0 \delta u - B_*  \right|.
\end{equation}
The expression was simplified by assuming \(\delta u \gg 1 / \gamma^2\), which holds true by a significant margin in the regimes of interest. It follows from \cref{E-effective-2} that the impact of the laser field is negligible if
\begin{equation} \label{cond-1}
    |B_*|/E_0 \gg \delta u.
\end{equation}
Under this condition, we have
\begin{equation} \label{E-tp}
    {\cal{F}} \approx \left| B_*  \right|.
\end{equation} 
It is important to note that, although the laser fields do not appear in the final expression for ${\cal{F}}$, they are still much stronger than the plasma magnetic field.

The impact of the laser fields can increase as the electron moves away from the turning point, primarily due to the increase in \(|\theta|\), which reaches its maximum value when the electron reaches the axis of the magnetic filament. On this axis, the plasma magnetic field \(B^{pl}\) is zero. Using the expression from \cref{E-effective}, we find that 
\begin{equation} \label{E-effective-3}
    {\cal{F}} \leq \left| \cos \theta_0 - 1/u \right| E_0,
\end{equation}
where \(\theta_0\) is the angle between \(\bm{p}\) and the axis of the filament. Here, we again assume that \(\delta u \gg 1 / \gamma^2\). For small angles (\(|\theta_0| \ll 1\)), we can approximate \(\cos \theta_0 \approx 1 - \theta_0^2 /2\), simplifying the expression to
\begin{equation} \label{E-effective-4}
    {\cal{F}} \leq \left| \delta u - \theta_0^2/2 \right| E_0.
\end{equation}
From \cref{fig: efficient_particle}c, we infer that \({\cal{F}}\) on the axis of the filament is much smaller than \({\cal{F}}\) at the turning points. Comparing \cref{E-effective-4} with \cref{E-tp}, we conclude that this requires
\begin{equation} \label{cond-2}
    |B_*|/E_0 \gg \left| \delta u - \theta_0^2/2 \right|.
\end{equation}
One immediate takeaway from this result is that $\theta_0$ must be sufficiently small. 

Equations (\ref{cond-1}) and (\ref{cond-2}) establish conditions that ensure \(B^{pl}\) dominates the emission process during DLA. These constraints can be interpreted as conditions on \(\alpha\) and \(\delta u\), which we will explore further at the end of this section through a corresponding parameter scan. For now, it is sufficient to note that in the efficient DLA example shown in \cref{fig: efficient_particle}, both conditions are satisfied. Near the highest peak of \(\chi^2\), we have \(|y|/\lambda_0 \approx 2\), leading to \(|B_*|/E_0 \approx 7.6/40 \approx 0.2\). Since \(\delta u = 0.01\), it follows that \(|B_*|/E_0 \gg \delta u\). The largest angle near the highest peak of \(\chi^2\) is \(\theta_0 \approx 10^\circ\), or equivalently, \(\theta_0 \approx 0.17\). This gives \(\theta_0^2/2 \approx 1.5 \times 10^{-3}\) and \(\theta_0^2/2 - \delta u \approx 5 \times 10^{-3}\). Consequently, \(|B_*|/E_0 \gg \left| \delta u - \theta_0^2/2 \right|\).

To demonstrate the critical role of the efficient DLA regime in achieving a single-peak emission profile, we performed an additional test-particle calculation. The parameters for this calculation are detailed in the `Inefficient DLA' column of \cref{table: efficient and inefficient}, with the only difference being a lower normalized current density: \(\alpha\) is now 1.0, compared to 12.0 in the efficient DLA regime. The resulting trajectory is shown in \cref{fig: inefficient_particle}a. In this scenario, the electron fails to achieve frequency matching, which prevents prolonged energy gain. This is evident from the alternating yellow and dark green segments along the trajectory, indicating energy gain and loss, respectively. As a result, the peak \(\gamma\) value does not exceed 600, in stark contrast to the efficient DLA case, where \(\gamma\) reached nearly 3000. We refer to this scenario as inefficient DLA. A key distinction from the efficient DLA regime is that the laser fields exert a much more prominent influence on the electron dynamics, as inferred from the significant fluctuations in the 
$\gamma$-factor shown in \cref{fig: inefficient_particle}b. 

As with the efficient DLA case, we assess the impact of the laser fields on the inefficient DLA case by comparing its \(\rchi\), \(\rchi_f\), and \(\rchi_L\) as shown in \cref{fig: inefficient_particle}c. In contrast to the efficient DLA case, \(\rchi^2\) and \(\rchi_f^2\) diverge significantly, with \(\rchi^2\) peaks being approximately twice as high as those of \(\rchi_f^2\). This discrepancy highlights the prominent influence of the laser fields. The primary reason for this change is the reduced value of \(\alpha\) compared to the efficient DLA case. Near the highest peak of \(\chi^2\), we again have \(|y|/\lambda_0 \approx 2\). However, due to the normalized current density being weaker by more than an order of magnitude, \(|B_*|/E_0\) is now approximately \(1.7 \times 10^{-2}\). Consequently, the condition given by \cref{cond-1} is no longer satisfied, with \(|B_*|/E_0 \sim \delta u = 0.01\). This further confirms that the laser field's effect on emission in this case is not negligible.

Although the laser does not entirely dominate the electron dynamics in this scenario, its increased influence compared to the efficient DLA case is enough to alter the photon emission pattern. \Cref{fig: dist matlab}a shows the angular probability distribution \({\cal P}_{\theta}\) and the angular distribution of the emitted energy \({\cal P}_{\gamma}\) for the trajectory presented in \cref{fig: inefficient_particle}a. The distribution \({\cal P}_{\theta}\) again exhibits two primary peaks. However, \({\cal P}_{\gamma}\) now also features two distinct peaks rather than one. They are located at \(\theta \approx 9^{\circ}\) and \(\theta \approx -9^{\circ}\). These peaks are not as sharp as those observed in \cref{fig: DLAfilamanet_dist_MATLAB}a for a free electron irradiated by the laser due to the influence of the magnetic filament. 

\rc{In conclusion, we introduced the concept of efficient DLA — a regime in which the electron steadily accumulates energy over multiple laser cycles. A distinctive feature of this regime, resulting from the prolonged energy gain, is the relatively small modulations in \(\gamma\). We found that this characteristic enables the formation of a single-peaked emission profile, in contrast to inefficient DLA — a regime defined by much stronger modulations of the \(\gamma\)-factor due to lower energy gain.}

%In conclusion, we introduced a concept of efficient DLA - a regime where the electron steadily accumulates energy over multiple laser periods. Its distinctive feature, resulting from the the prolonged energy gain, are the relatively small modulations in \(\gamma\). We found that it is this feature that enables the formation of a single-lobed emission profile, compared to the case of inefficient DLA where the \(\gamma\)-factor experiences much stronger modulations. %These modulations appear small primarily because of the significant overall energy gain achieved throughout the DLA process. 

%*****************************

\section{Parameter scan and analysis of emission efficiency} \label{sec: parameter scan}

In \cref{sec: test electron model}, we analyzed two examples of electron dynamics and demonstrated that an electron in the efficient DLA regime produces a distinctly different photon emission profile compared to an electron in the inefficient DLA regime. In this section, we conduct a parameter scan to confirm that this trend is general. Additionally, we examine the emission efficiency, showing that efficient DLA is directly associated with significantly higher photon emission. %, with the electron emitting orders of magnitude more energy than in the inefficient DLA regime.

\begin{figure}
    \centering
    \includegraphics[width=.75\linewidth]{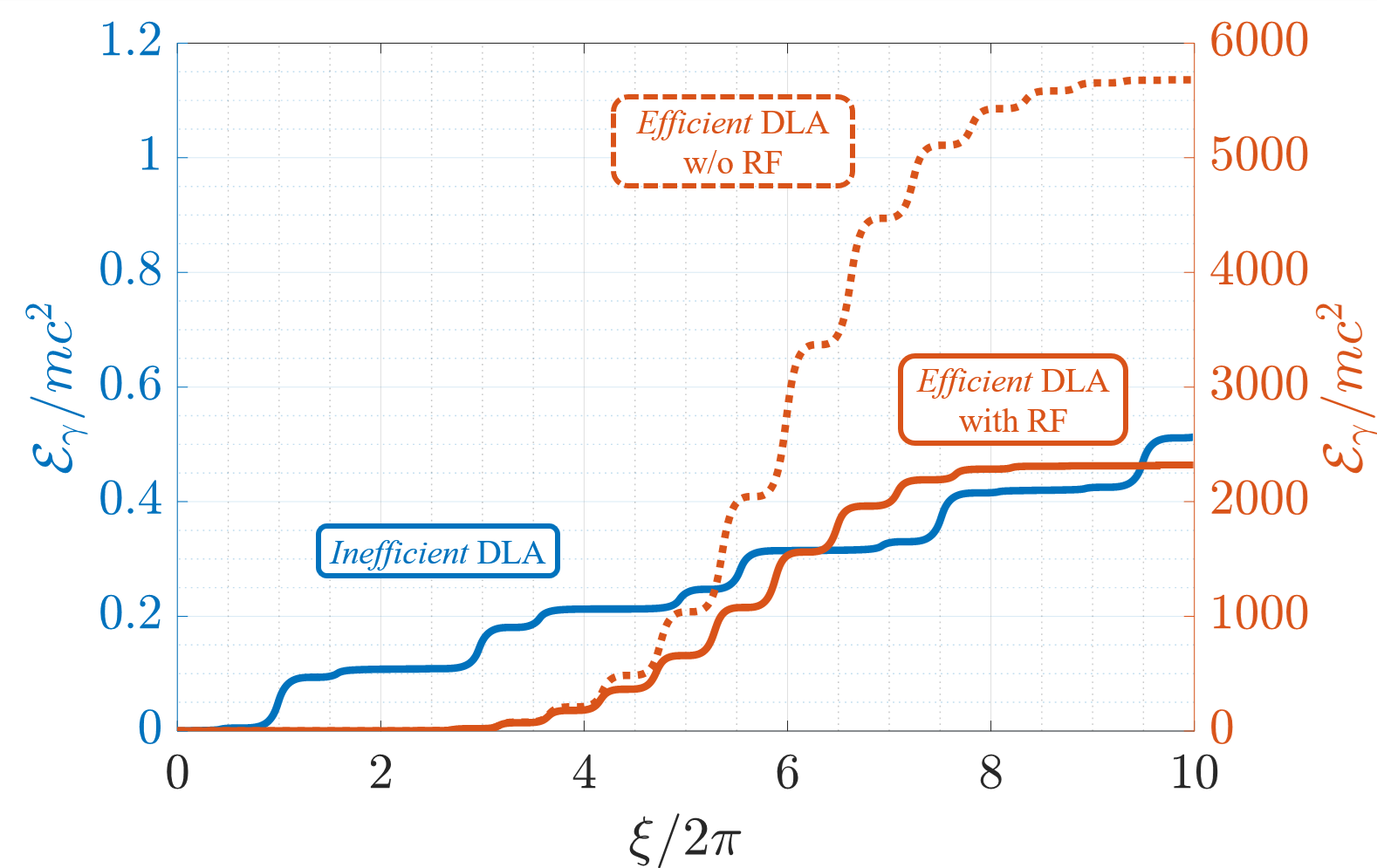}
    \caption{Comparison of the emitted energy in the examples shown in \cref{fig: inefficient_particle} and \cref{fig: efficient_particle}, representing the \textit{inefficient} and \textit{efficient} DLA regimes. The accumulated emitted energy, \(\mathcal{E}_{\gamma}\), is normalized by \(mc^2\) and plotted as a function of the laser phase. The \textit{inefficient} DLA regime from \cref{fig: inefficient_particle} is depicted by the solid blue curve (with the blue scale on the left), while the \textit{efficient} DLA regime from \cref{fig: efficient_particle} is depicted by the solid red curve (with the red scale on the right). For comparison, the dotted red curve represents the emission in the \textit{efficient} DLA regime if the radiation friction force is neglected. }
    \label{fig: ba accum energy MATLAB}
\end{figure}

We begin this section by comparing photon emission in the examples shown in \cref{fig: inefficient_particle} and \cref{fig: efficient_particle}, which represent the \textit{inefficient} and \textit{efficient} DLA regimes. \Cref{fig: ba accum energy MATLAB} presents the accumulated emitted energy, \(\mathcal{E}_{\gamma}\),  as a function of the laser phase to facilitate correlation between changes in emission and variations in $\gamma$ and $\chi^2$ observed  in \cref{fig: inefficient_particle} and \cref{fig: efficient_particle}. The emitted energy is normalized by \(mc^2\). In the inefficient DLA regime, the electron emits approximately $0.5 m c^2$ after 10 laser oscillations, which is significantly less than its peak energy ($\sim 500 mc^2$). On average, the emission occurs at roughly the same rate. In contrast, in the efficient DLA regime, the emission rate increases significantly, particularly between $\xi/2\pi \approx 2$ and $\xi/2\pi \approx 6$, as result of the rise in $\chi^2$ seen in \cref{fig: efficient_particle}c. This increase enables the electron to emit about 5000 times more energy. The total emitted energy ($\sim 2200 mc^2$) is only slightly less than the peak electron energy ($\sim 2900 mc^2$).

The increase in the emission rate in the efficient DLA regime is directly linked to the increase in electron energy. This is evident by comparing \cref{fig: efficient_particle}c, which shows $\chi^2$, and \cref{fig: efficient_particle}b, which shows $\gamma$. Additionally, the fact that efficient DLA is achieved at a higher value of $\alpha$ in the two examples further amplifies the difference in total emission. The emission becomes so intense that it begins to influence the electron dynamics at approximately $\xi/2 \pi > 4$. In this regime, the inclusion of the radiation friction force is essential for accurately calculating the electron dynamics. Neglecting this force can lead to significant overestimation of the emitted energy. This effect is illustrated in \cref{fig: ba accum energy MATLAB}, where the emitted energy is more than double in a calculation that omits the radiation friction force.

%, representing the feedback of themission 

%The total energy emitted by a single electron in the \textit{inefficient} DLA regime from \cref{fig: inefficient_particle} (blue curve, with the blue scale on the left) and the \textit{efficient} DLA regime from \cref{fig: efficient_particle} (red curve, with the red scale on the right) as a function of the laser phase. The emitted energy \(\mathcal{E}_{\gamma}\) is normalized by \(mc^2\).  

The key insight from the two examples obtained using test-electron calculations is that an increase in DLA efficiency is associated with the transition from a double-peaked to a single-peaked emission profile, along with enhanced emission. To confirm the generality of this trend, we conducted a parameter scan by varying \(\alpha\) and \(\delta u \equiv (v_{ph} - c)/c\). Additional details of this scan are provided in \cref{table: test electron scan}. All other parameters, including the laser amplitude and the initial electron conditions, were kept consistent with those listed in \cref{table: efficient and inefficient}. For simplicity, the electron’s initial transverse displacement was held constant for each set of \(\alpha\) and \(\delta u\). While this deliberate simplification limits the scope of the scan, it is sufficient to establish the general connection between DLA efficiency, emission profiles, and enhanced emission.

\begin{table}[b!]
\footnotesize
\centering
\begin{tabular}{ | p{6cm} | p{6cm}| }
\hline
\multicolumn{2}{|c|}{Parameter scan using test-electron model} \\
\hline
Normalized current density range  & $0.2 \leq \alpha \leq 16.0$ \\
\hline
Number of elements along $\alpha$  & 300 (evenly spaced) \\
\hline
% Normalized phase-velocity range  & $1.005 \leq v_{ph}/c \leq 1.03$ \\
% \hline
Range of relative superluminosity & $0.005 \leq \delta u \leq 0.03$ \\
\hline
Number of elements along $\delta u$  & 300 (evenly spaced) \\
\hline
Duration of each run & 500 laser periods \\
\hline
\end{tabular}
\caption{Details of the parameter scan presented in \cref{fig: scan}, \cref{fig: chi_ratio}, and \cref{fig: emission compare}. All other parameters are the same as those in \cref{table: efficient and inefficient}.}
\label{table: test electron scan}
\end{table}

\begin{figure*}
    \centering
    \includegraphics[width=1\textwidth]{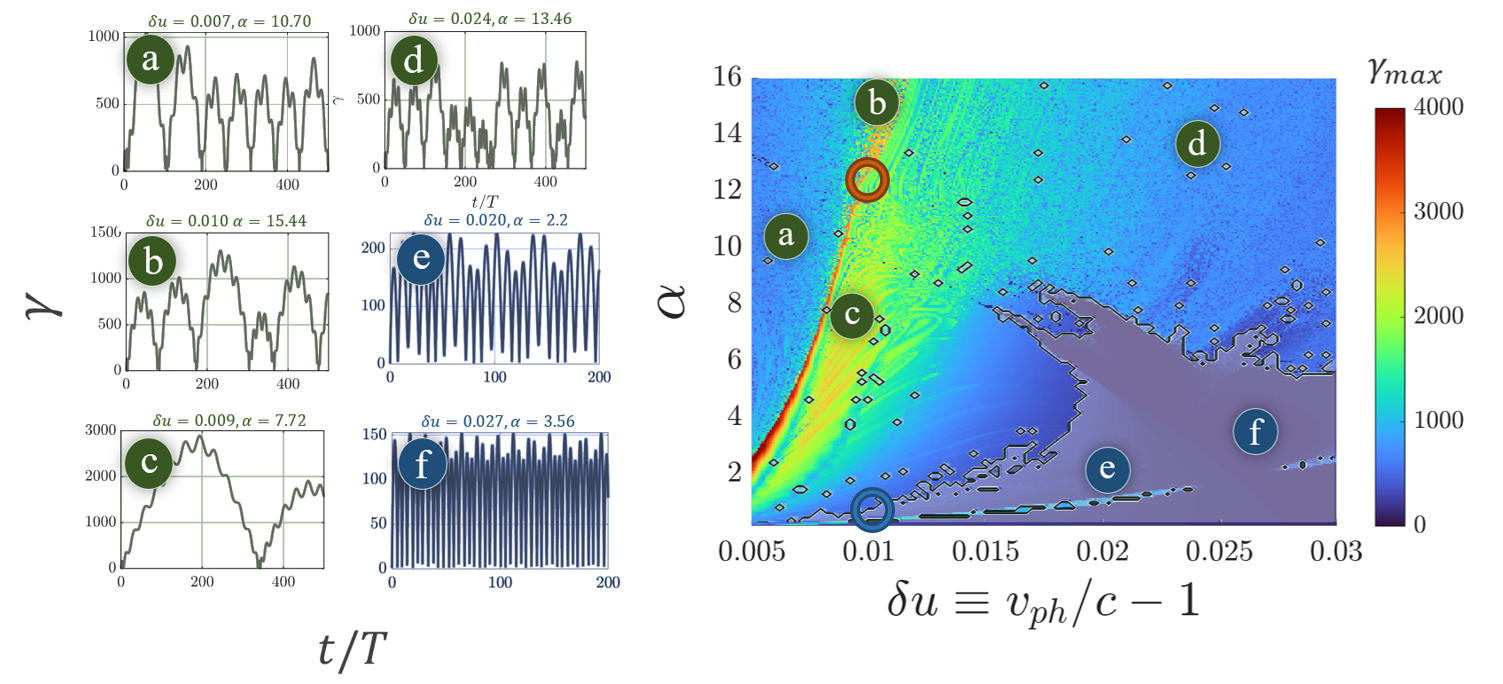}
    \caption{The maximum relativistic factor, \(\gamma_{\max}\), achieved by an electron irradiated by a plane wave in a magnetic filament, as a function of the normalized current density \(\alpha\) and the relative degree of superluminosity \(\delta u \equiv (v_{ph} - c)/c\). Regions with a gray overlay correspond to cases with double-peaked emission profiles, while other regions exhibit a single-peaked profile. All parameters except \(\alpha\) and \(\delta u\) are fixed, consistent with those listed in \cref{table: efficient and inefficient}. Additional details of the scan are provided in \cref{table: test electron scan}. Panels  (a) through (f) show the time evolution of \(\gamma\) for six selected test-electron calculations, corresponding to the parameter sets marked with the same letters on the scan. The red open circle indicates the efficient DLA regime from \cref{fig: efficient_particle}, while the blue open circle marks the inefficient DLA regime from \cref{fig: inefficient_particle}.    }
    \label{fig: scan}
\end{figure*}

\Cref{fig: scan} shows the maximum relativistic factor, $\gamma_{\max}$, achieved by the electron for each combination of \(\alpha\) and \(\delta u\).  Generally, high \(\gamma_{\max}\) values indicate that the electron has experienced efficient DLA. For each parameter set, we also calculated the angular distribution of emitted energy, \({\cal P}_{\gamma}\), and determined whether the distribution featured one or two peaks. The algorithm used for this analysis is detailed in \ref{appendix: algorithm}, which includes additional examples of emission profiles. Regions with a gray overlay correspond to cases with double-peaked emission profiles, while other regions exhibit a single-peaked profile. The scan broadly confirms that entering the efficient DLA regime and achieving high values of \(\gamma_{\max}\) is correlated with having a single-peaked emission profile.

%The key insight from the two examples obtained using test-electron calculations is that a change (increase) in DLA efficiency is linked to the change from double- and single-lobed emission profile and enhanced emission. To confirm that this trend is general, we conducted a parameter scan, as shown in \cref{fig: scan}, by varying \(\alpha\) and \(\delta u \equiv (v_{ph} -c)/c\). Additional details of this scan can be found in \cref{table: test electron scan}. All other parameters, including the laser amplitude and the initial electron conditions, were kept consistent with those listed in \cref{table: efficient and inefficient}. The color scale in the figure represents the maximum relativistic factor \(\gamma_{\max}\) achieved by the electron for each combination of \(\alpha\) and \(\delta u\). Generally, high \(\gamma_{\max}\) values indicate that the electron has experienced efficient DLA. For each parameter set, we also calculated the angular distribution of emitted energy \({\cal P}_{\gamma}\) and identified whether the distribution featured one or two lobes. Regions with a gray overlay correspond to cases with double-lobed emission profiles, while other regions exhibit a single-lobed profile. The scan broadly confirms that entering the efficient DLA regime and achieving high values of \(\gamma_{\max}\) is correlated with having a single-lobed emission profile.

\begin{figure*}
    \centering
    \includegraphics[width=1\textwidth]{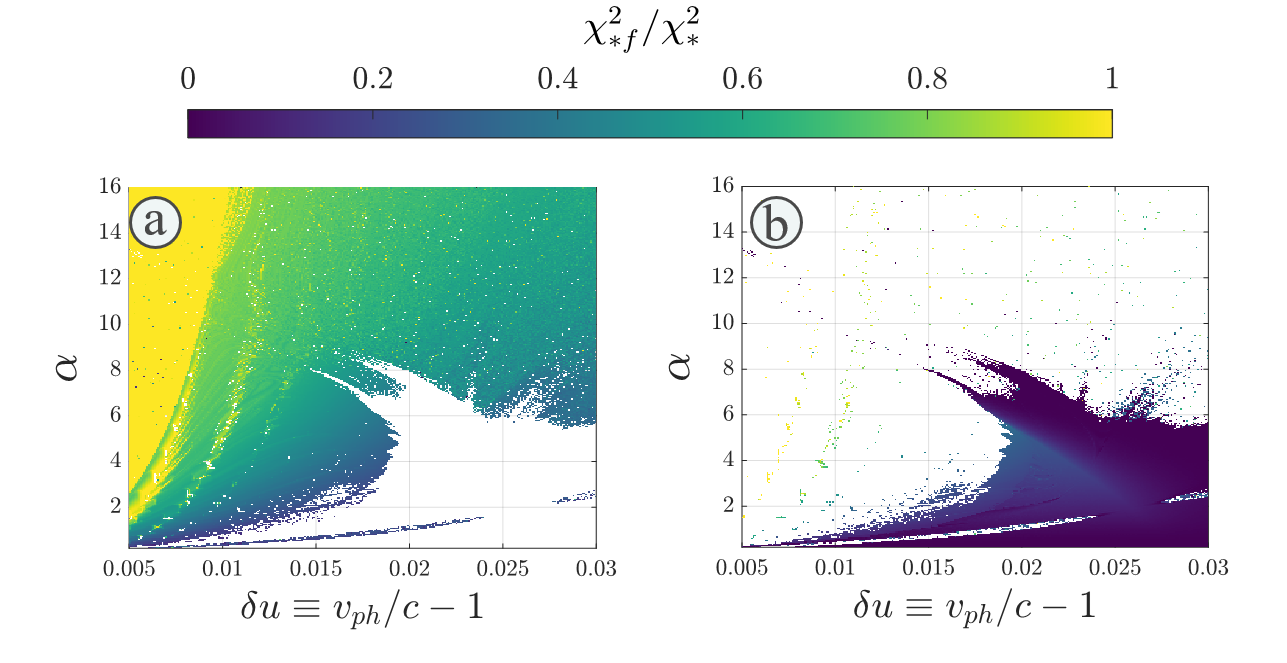}
    \caption{The fraction of the power the electron would emit if it only experienced the field of the filament, calculated at the most prolific turning point. The parameter scan is the same as in \cref{fig: scan}. The definitions of $\rchi_{*}$ and $\rchi_{*f}$ are provided in the main text. (a) Only the values corresponding to single-peaked emission are shown. (b) Only the values corresponding to double-peaked emission are shown. This is the region marked with the gray overlay in \cref{fig: scan}.}
    \label{fig: chi_ratio}
\end{figure*}

\Cref{fig: chi_ratio} demonstrates that the single-peaked emission observed in the parameter scan is primarily driven by the plasma magnetic field. The figure compares two quantities. One is \(\chi_{*}^2\), which represents the highest value of \(\chi^2\) along the electron trajectory. As discussed earlier, \(\chi^2\) reaches its peaks at turning points, so \(\chi_{*}^2\) corresponds to the highest peak. The other quantity is \(\chi_{*f}^2\), the value of \(\chi^2\) at the same peak, but calculated using only the plasma magnetic field. The ratio $\chi_{*f}^2/\chi_*^2$ represents the fraction of the power emitted at the peak that the electron would emit if it only experienced the field of the filament. \Cref{fig: chi_ratio}a shows \(\chi_{*f}^2/\chi_*^2\) for parameter sets that produce a single-peaked emission profile, while \cref{fig: chi_ratio}b presents the same ratio for parameter sets that produce a double-peaked emission profile. In \cref{fig: chi_ratio}b, \(\chi_{*f}^2/\chi_*^2\) is low, which means that the laser fields significantly influence the emission in the two-peaked regime, which is also the regime of inefficient DLA. In contrast, \cref{fig: chi_ratio}a shows significantly higher values of \(\chi_{*f}^2/\chi_*^2\). As expected from the conditions given by \cref{cond-1}, \(\chi_{*f}^2/\chi_*^2\) increases with higher \(\alpha\) and lower \(\delta u\). Notably, \(\chi_{*f}^2/\chi_*^2\) approaches unity for all parameter sets in \cref{fig: scan} where \(\gamma_{\max}\) reaches its highest values and exhibits a single-peaked emission profile.

\begin{figure} [h!]
     \centering
     \includegraphics[width=0.6\linewidth]{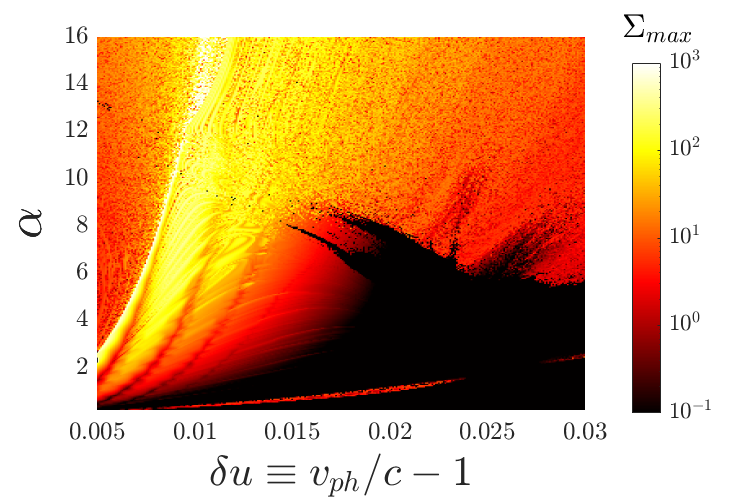}
     \caption{The maximum emission per laser oscillation, \(\Sigma_{\max}\), as a function of the normalized current density \(\alpha\) and the relative degree of superluminosity, \(\delta u \equiv (v_{ph} - c)/c\). \(\Sigma_{\max}\) is normalized to \(mc^2\) and is determined by analyzing the entire electron trajectory for each set of parameters (\(\alpha\) and \(\delta u\)) and identifying the highest emission within a single laser oscillation (\(\Delta \xi = 2\pi\)).}
     \label{fig: emission compare}
 \end{figure}

Finally, \cref{fig: emission compare} presents the photon emission for the same parameter scan, showing the maximum emission per laser oscillation, \(\Sigma_{\max}\), normalized to \(mc^2\). This quantity is determined by analyzing the entire electron trajectory for each set of parameters (\(\alpha\) and \(\delta u\)) and identifying the highest emission within a single laser oscillation (\(\Delta \xi = 2\pi\)). The dynamic range of \(\Sigma_{\max}\) spans four orders of magnitude, necessitating the use of a logarithmic scale for the color-coding. The scan broadly confirms that entering the efficient DLA regime and achieving high \(\gamma_{\max}\) values leads to a dramatic increase in emitted energy. This variation occurs even at a fixed \(\alpha\), which corresponds to a fixed magnetic field configuration that drives the emission. For instance, at \(\alpha = 2\), \(\Sigma_{\max}\) increases from approximately \(10^{-1}\) to \(10^3\) as the regime transitions from inefficient DLA to efficient DLA.

To conclude this section, it is important to emphasize the key characteristic of the efficient DLA regime that enables the formation of a single-peaked emission profile: the small relative modulations in \(\gamma\) during a single betatron oscillation. These modulations appear small primarily because of the significant overall energy gain achieved throughout the DLA process. \Cref{fig: scan} illustrates this with six examples, shown in the panels on the left side of the figure. Each panel displays \(\gamma\) as a function of time \(t\) for different combinations of \(\alpha\) and \(\delta u\), labeled with corresponding letters that are also marked on the scan. In panels (a) through (d), the parameters result in a single-peaked emission profile, directly correlated with achieving high \(\gamma_{\max}\) values and exhibiting only minor modulations in \(\gamma\). In contrast, the parameters in panels (e) and (f) lead to a double-peaked profile, where the lower energy gain causes more pronounced fluctuations in \(\gamma\).

%The accumulated emitted $\gamma$-ray energy $\mathcal{E}_{\gamma}$ normalized by $mc^2$.  The blue curve represents the electron emission from the \textit{inefficient} DLA regime in \cref{fig: inefficient_particle} while the red curve represents the electron emission from the \textit{efficient} DLA regime in \cref{fig: efficient_particle}. 

%%%%%%%%%%%%%%%%%%%%%%%%%%%%%%%%%%%%%%%%%%%%%%%%%%%%%%%%%%%%%%%%%%%%%%%%%%%%%%%%%%%%%%%%%%%%%%%%%%%%%%%%%%

\section{Summary and discussion} \label{Sec: summary and discussion}

In this study, we explored the conditions under which direct laser acceleration (DLA) of electrons leads to distinct photon emission profiles, with a particular focus on identifying the mechanisms that produce single-lobed versus double-lobed angular distributions of emitted \(\gamma\)-rays. Our analysis reveals that the efficiency of DLA plays a crucial role in shaping the emission profile. 

Electrons that gain and lose a significant portion of their energy over a single laser cycle — characteristic of inefficient DLA — tend to emit \(\gamma\)-rays in a double-peaked angular distribution. This emission pattern is a direct consequence of the electron dynamics, where the laser fields have a prominent influence on the emission process.

In the efficient DLA regime, electrons steadily accumulate energy over multiple laser cycles, and the emission is dominated by the quasi-static azimuthal magnetic field generated in the plasma by the laser. This regime produces a single-peaked angular distribution of emitted \(\gamma\)-rays, even though the electrons themselves exhibit a two-peaked angular distribution. 

The single-peaked emission arises primarily from the turning points of the electron trajectory, where the influence of the plasma magnetic field is most significant. In this regime, electrons achieve much higher energies and, as result, emit orders of magnitude more energy due to the fact that the emitted power scales as $\gamma^2$. For instance, in the presented scan, the maximum emission during a single laser oscillation can be as high as \(10^3 mc^2\) for the electron experiencing efficient DLA.

We further demonstrated, using fully self-consistent PIC simulations, that the single-peaked emission regime can be achieved by lowering the electron density in the target. The lower density target considered in this work creates favorable conditions for some electrons to enter the efficient DLA regime. As these electrons achieve much higher values of \(\gamma\) and \(\chi^2\), they begin to dominate the emission, transforming its profile from a two-peaked to a single-peaked shape. While electrons experiencing inefficient DLA are still present, their emission is overpowered by that of the electrons in the efficient DLA regime.

Our study offers critical insights into the mechanisms driving \(\gamma\)-ray emission in laser-plasma interactions, highlighting potential pathways for optimizing laser-driven \(\gamma\)-ray sources. According to our findings, these optimizations must be closely tied to creating conditions where electrons can enter the regime of efficient DLA, as its impact on the \(\gamma\)-ray emission profile has often been overlooked. This focus is essential for improving the practical utility of \(\gamma\)-ray beams in applications requiring high-intensity, well-collimated sources.

%In this work, we focused specifically on electron acceleration by a linearly polarized laser beam. This is because most experimental research on DLA employs linearly polarized lasers. This is largely because linear polarization is the native configuration for conventional laser systems. For this reason, we have limited our focus to linearly polarized lasers, which represent the majority of experimental capabilities and experimental research in this area.

%In this work, we focused on electron acceleration by a linearly polarized laser beam, as this is the native configuration for conventional laser systems and is widely used in experimental research on DLA. By concentrating on linearly polarized lasers, we align our study with the majority of experimental capabilities and research efforts in this area. However, other laser polarizations, such as circular and radial, are also being explored in the context of ultra-high-intensity laser-plasma experiments. Moreover, there exist additional options, such a structured light with twisted wavefronts. It is not possible to directly generalize our research, as one first needs to understand how these polarizations generate magnetic fields in the plasma and how they accelerate electrons. Nevertehless, our results that show qualitative changes suggest that examining these options may be warranted.

\rc{In this work, we focused on electron acceleration by a linearly polarized laser beam, as it is the native polarization for conventional high-power laser systems and widely used in experimental research on DLA~\cite{Hussein.NJP.2021.Towards, Tang.NJP.2024.Focusing, Willingale.NJP.2018.Longitudinal, Cohen.SA.2024.Undepleted}. However, alternative polarizations, such as circular and radial, as well as structured light with twisted wavefronts~\cite{Shi2024}, are beginning to emerge in ultra-high-intensity laser-plasma experiments. Generalizing our findings to these configurations is not straightforward, as their effects on plasma magnetic fields and electron acceleration must first be understood. Nonetheless, the significant improvements we discovered for linear polarization suggest that exploring these alternatives could be highly worthwhile.}

\rc{Our work underscores the necessity of including the force of radiation friction when considering efficient DLA, even at relatively modest values of \(a_0\). The conventional expectation that radiation friction becomes significant only when \(a_0\) reaches the hundreds—rather than around 40, as in our study—arises from the assumption that the electron \(\gamma\)-factor, which determines the magnitude of the force, is roughly equal to \(a_0\). While this assumption may hold for the bulk of electrons, it does not apply to electrons experiencing efficient DLA. For example, in the presented scan, the value of \(\gamma\) for such electrons is two orders of magnitude higher.}

\rc{Although a detailed investigation of the role of radiation friction is beyond the scope of this study, we highlight a few key features. Radiation friction generally reduces the energy gain for electrons that enter the regime of efficient DLA. However, it can also have a counterintuitive effect: enabling electrons that would otherwise fail to enter the efficient DLA regime to do so by altering a parameter that would otherwise be a conserved integral of motion~\cite{Gong.SR.2019.Radiation, Yeh.NJP.2021.Friction}. These effects highlight the complex interplay between radiation friction and the DLA regime. In \ref{appendix: Role RR}, we present plots from the parameter scan discussed in \cref{sec: parameter scan}, recalculated without including radiation friction. These plots confirm that some electrons in our scan enter the efficient DLA regime and produce a single-peaked emission profile only when radiation friction is included. However, these are not the most energetic cases.}

%Our work also underscores the necessity of including the force of radiation friction when considering efficient DLA, even at relatively modest values of \(a_0\). The conventional expectation that radiation friction becomes significant only when \(a_0\) reaches the hundreds—rather than around 40, as in our study—arises from the assumption that the electron \(\gamma\)-factor, which determines the magnitude of the force, is roughly equal to \(a_0\). While this assumption may hold for the bulk of electrons, it does not apply to electrons experiencing efficient DLA. For instance, in the presented scan, the value of \(\gamma\) is two orders of magnitude higher. 

%A study of the role played by the radiation friction goes beyond the scope of our study, so want to just point out just several important features.

%The radiation friction tends to slow down the energy gain for the electrons that enter the regime of efficient DLA.  However, the electrons that otherwise would not enter this regime, can become eligible to enter the regime of efficient DLA due to the radiation friction. In this case, the radiation friction changes the quantity that determines the energy gain and that would be an integral of motion in its absence and . 

%%%%%%%%%%%%%%%%%%%%%%%%%%%%%%%%%%%%%%%%%%%%%%%%%%%%%%%%%%%%%%%%%%%%%%%%%%%%%%%%%%%%%%%%%%%%%%%%%%%%%%%%%%

\section{Acknowledgements}

This research was supported by the National Science Foundation–Czech Science Foundation Partnership (NSF Grant No. PHY-2206777 and GACR project No. 22-42890L) and by the United States-Israel Binational Science Foundation (BSF Grant No. 2022322).  We acknowledge the Texas Advanced Computing Center (TACC) at The University of Texas at Austin for providing computational resources through the Stampede2 and Stampede3 supercomputers that have contributed to the research results reported in this paper. The TACC resources were allocated through project number PHY190034 from the Advanced Cyberinfrastructure Coordination Ecosystem: Services and Support (ACCESS) program, which is supported by National Science Foundation grants 2138259, 2138286, 2138307, 2137603, and 2138296.

%%%%%%%%%%%%%%%%%%%%%%%%%%%%%%%%%%%%%%%%%%%%%%%%%%%%%%%%%%%%%%%%%%%%%%%%%%%%%%%%%%%%%%%%%%%%%%%%%%%%%%%%%%%%%%%%
\newcommand{\newblock}{}
\bibliographystyle{apsrev4-1}
\bibliography{Collections}

%%%%%%%%%%%%%%%%%%%%%%%%%%%%%%%%%%%%%%%%%%%%%%%%%%%%%%%%%%%%%%%%%%%%%%%%%%%%%%%%%%%%%%%%%%%%%%%%%%%%%%%%%%%%%%%%
\newpage
\appendix

%%%%%%%%%%%%%%%%%%%%%%%%%%%%%%%%%%%%%%%%%%%%%%%%%%%%%%%%%%%%%%%%%%%%%%%%%%%%%%%%%%%%%%%%%%%%%%%%%%%%%%%%%%%%%%%%

\section{Angular distribution of photon emission in 3D PIC simulations for different photon energy ranges} \label{Appendix A}

In the main text, \cref{fig: pic photons} shows the accumulated emitted energy per steradian, calculated by projecting all photons with \(\varepsilon_{\gamma} > 10\) keV onto a sphere. Here, we present additional post-processed data from the same two 3D PIC simulations, using different photon energy ranges to calculate the accumulated emitted energy.

\begin{figure} [h!]
    \centering
    \includegraphics[width=0.8\linewidth]{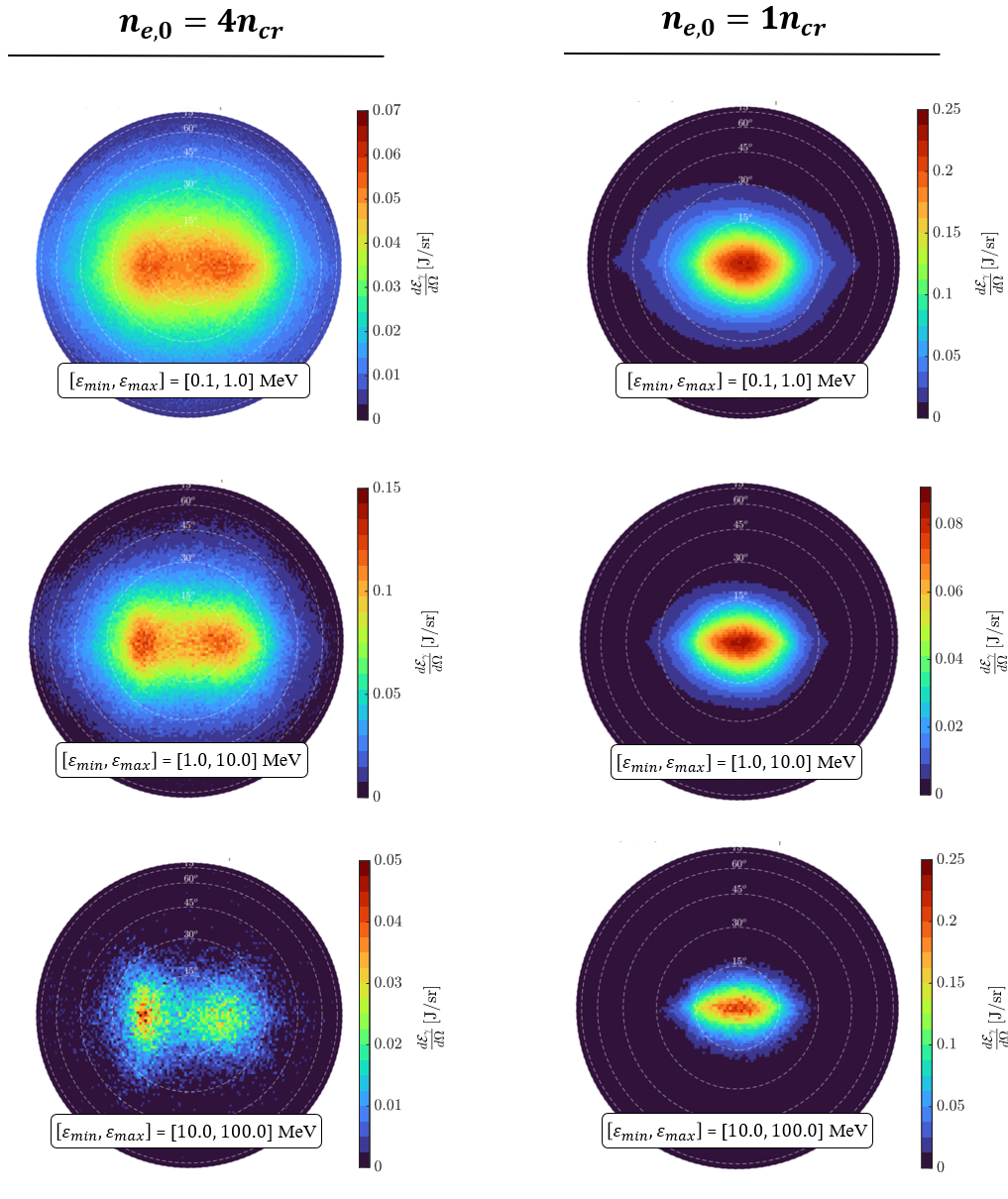}
    \caption{Angular distribution of the emitted energy, $d {\cal{E}}_{\gamma} / d \Omega$, from simulations with $n_{e,0} = 4n_{cr}$ (left column) and $n_{e,0} = 1n_{cr}$ (right column) for photons with $\varepsilon_{\gamma}$ in the range between $\varepsilon_{\min}$ and $\varepsilon_{\max}$. In the upper row, $\varepsilon_{\min} =0.1$~MeV and $\varepsilon_{\max} = 1$~MeV; in the middle row, $\varepsilon_{\min} =1$~MeV and $\varepsilon_{\max} = 10$~MeV; and in the bottom row, $\varepsilon_{\min} =10$~MeV and $\varepsilon_{\max} = 100$~MeV.}
    \label{fig: energy rangs photon ang dist}
\end{figure}

In Figure~\ref{fig: energy rangs photon ang dist}, the left column corresponds to the simulation with \(n_{e,0} = 4n_{cr}\), and the right column corresponds to the simulation with \(n_{e,0} = 1n_{cr}\). The rows represent different photon energy ranges: 100 keV \(> \varepsilon_{\gamma} > 1.0 \) MeV (top row), 1.0 MeV \(> \varepsilon_{\gamma} > 10.0 \) MeV (middle row), and 10.0 MeV \(> \varepsilon_{\gamma} > 100.0 \) MeV (bottom row). In the simulation with \(n_{e,0} = 4n_{cr}\), the emission pattern consistently exhibits a two-peaked structure in the laser polarization plane. In contrast, the simulation with \(n_{e,0} = 1n_{cr}\) produces an emission pattern with a single peak aligned with the laser propagation direction. The accumulated emitted energy for each range is 0.14 J, 0.12 J, and 0.011 J for \(n_{e,0} = 4n_{cr}\) and 0.11 J, 0.20 J, and 0.030 J for \(n_{e,0} = 1n_{cr}\).

%%%%%%%%%%%%%%%%%%%%%%%%%%%%%%%%%%%%%%%%%%%%%%%%%%%%%%%%%%%%%%%%%%%%%%%%%%%%%%%%%%%%%%%%%%%%%%%%%%%%%%%%%%%%%%%%
\section{Determining single- versus double-peaked emission distributions in DLA test-electron trajectories } \label{appendix: algorithm}

In the parameter scan over \(\alpha\) and \(\delta u\) presented in \cref{fig: scan} and \cref{fig: chi_ratio}, we classified the angular distribution of emitted energy into two distinct categories: single-peaked and double-peaked profiles. This appendix details the procedure used to post-process the scan and determine whether a given set of parameters produces a single- or double-peaked emission profile.

\begin{algorithm} [b!]
	\caption{Pseudo-code for classifying the angular distribution of emitted energy into single- and double-peaked groups.} 
	\begin{algorithmic} 
            \State $\theta=$ the angle at each time snapshot 
            \State $P=$ the power of emitted $\gamma$-rays at each time snapshot 
            \newline
            \State $\mathcal{P}_{\gamma}$ = histogram(values $=\theta$, weights $=P$) 
            \State $\mathcal{P}_{\gamma}^{sm}$ = smooth($\mathcal{P}_{\gamma}$) 
            \State $\mathcal{D}_{\gamma}^{sm}$ = abs(diff(diff($\mathcal{P}_{\gamma}^{sm}$)))
            \State $\mathcal{D}_{\gamma,\text{right}}^{sm}=\mathcal{D}_{\gamma}^{sm};~ \mathcal{D}_{\gamma,\text{right}}^{sm}(\theta<\Theta) = \mathrm{NaN}$ 
            \State $\mathcal{D}_{\gamma,\text{left}}^{sm}=\mathcal{D}_{\gamma}^{sm};~ \mathcal{D}_{\gamma,\text{left}}^{sm}(\theta>-\Theta) = \mathrm{NaN}$ 
            \newline
            \State $\mathcal{P}_{\gamma}^\text{mid}$ = $\mathcal{P}_{\gamma}^{sm}(\theta=0)$
            \State $\mathcal{P}_{\gamma}^\text{right}$ = $\mathcal{P}_{\gamma}^{sm}(\mathcal{D}_{\gamma,\text{right}}^{sm}=\mathrm{max}(\mathcal{D}_{\gamma,\text{right}}^{sm}))$ 
            \State $\mathcal{P}_{\gamma}^\text{left}$ = $\mathcal{P}_{\gamma}^{sm}(\mathcal{D}_{\gamma,\text{left}}^{sm} =\mathrm{max}(\mathcal{D}_{\gamma,\text{left}}^{sm}))$ 
            \State $\mathcal{P}_{\gamma}^\text{sides} = (\mathcal{P}_{\gamma}^\text{right}+\mathcal{P}_{\gamma}^\text{left})/2$ 
            \newline
            \State $N_\text{peaks} = 1 + (\mathcal{P}_{\gamma}^\text{sides} > \mathcal{P}_{\gamma}^\text{mid})$
            \newline
	\end{algorithmic} 
        \label{algorithm}
\end{algorithm}

An algorithm is necessary for classifying the electron emission for two main reasons. First, some emission profiles are ambiguous. While most profiles are clearly single- or double-peaked, certain cases require an explicit criterion for classification. Second, the high-resolution scan involve a large number of trajectories. As outlined in Table~\ref{table: test electron scan}, the presented scan encompasses 90,000 test-electron simulations, varying \(\alpha\) and \(\delta u\). Given this volume, the classification of emission profiles must be fully automated, which necessitates the use of an algorithm.

The pseudo-code is detailed in Algorithm~\ref{algorithm}, and here we provide additional step-by-step explanations. The algorithm begins by calculating the angular distribution of emitted energy, \(\mathcal{P}_\gamma\), for a given electron trajectory. For ultra-relativistic electrons (\(\gamma \gg 1\)), the energy emitted at each angle is determined by weighting the time spent at that angle by the emission power. To reduce noise, \(\mathcal{P}_\gamma\) is smoothed using a moving average filter, producing \(\mathcal{P}_{\gamma}^\text{sm}\). The smoothed value at \(\theta = 0\) is assigned to \(\mathcal{P}_\gamma^\text{mid}\). Next, the algorithm identifies the points of highest second-order derivative of \(\mathcal{P}_{\gamma}^\text{sm}\) at angles \(\theta < -\Theta\) and \(\theta > \Theta\), assigning these values to \(\mathcal{P}_\gamma^\text{left}\) and \(\mathcal{P}_\gamma^\text{right}\), respectively. To avoid interference from high curvature near \(\theta = 0\), we exclude the region within \(|\theta| < \Theta\), where \(\Theta = 7^\circ\) in our scans. Finally, \(\mathcal{P}_{\gamma}^\text{sides}\) is computed as \((\mathcal{P}_\gamma^\text{left} + \mathcal{P}_\gamma^\text{right})/2\).  

The emission classification is performed by comparing \(\mathcal{P}_\gamma^\text{mid}\) with \(\mathcal{P}_{\gamma}^\text{sides}\). Distributions with \(\mathcal{P}_\gamma^\text{mid} < \mathcal{P}_{\gamma}^\text{sides}\) are classified as double-peaked, while those with \(\mathcal{P}_\gamma^\text{mid} > \mathcal{P}_{\gamma}^\text{sides}\) are classified as single-peaked. The value of \(\mathcal{P}_{\gamma}^\text{sides}\) is particularly meaningful for double-peaked distributions, quantifying the height of side peaks. For single-peaked distributions, the comparison acts more as a threshold, where \(\mathcal{P}_\gamma^\text{mid}/\mathcal{P}_{\gamma}^\text{sides} > 1\) indicates a single peak.

 \begin{figure} [t!]
     \centering
     \includegraphics[width=0.8\linewidth]{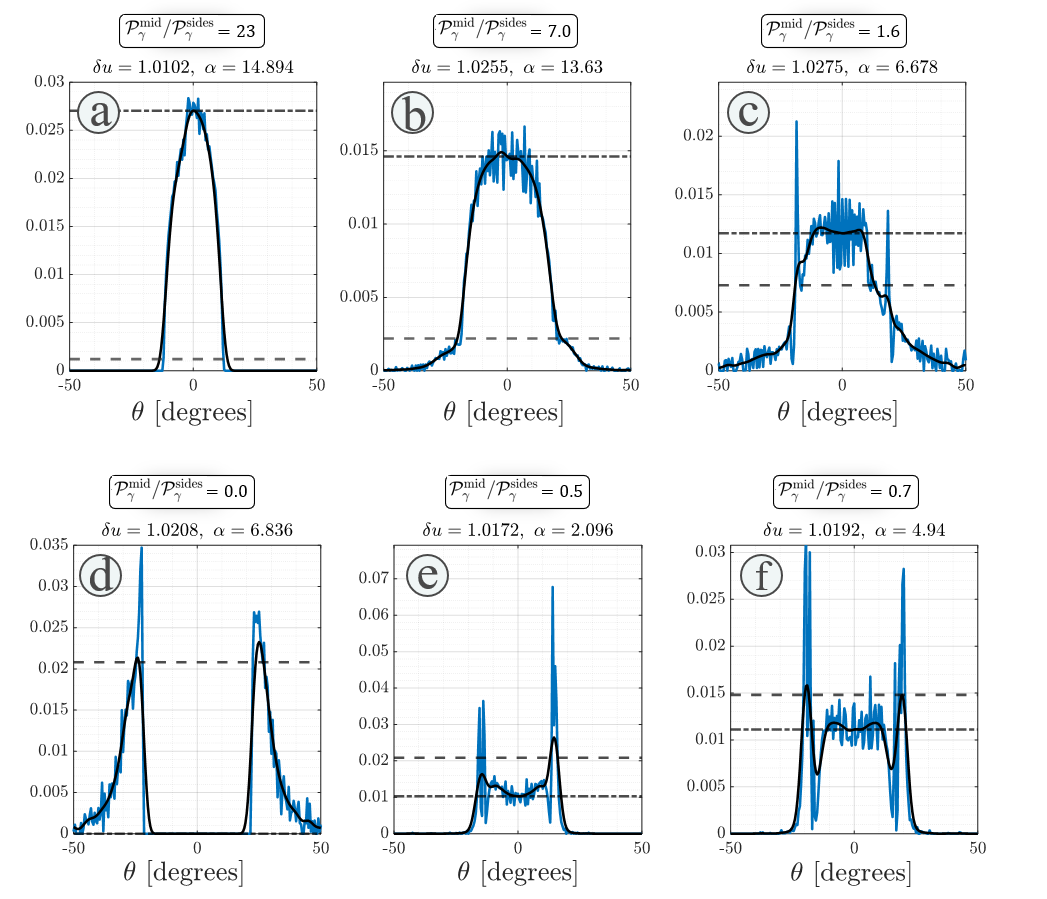}
     \caption{Example emission distributions from test-electron calculations. The dotted-dashed line indicates $\mathcal{P}_\gamma^\text{mid}$ and the dashed line represents $\mathcal{P}_\gamma^\text{sides}$.  The top row displays distributions classified as single-peaked, whereas the bottom row shows distributions classified as double-peaked.  }
     \label{fig: emission dist examples}
 \end{figure} 

%Example emission distributions from test-electron calculations. The smoothed distribution is in black where the dotted-dashed line is $\mathcal{P}_\gamma^\text{mid}$ and the dashed line is $\mathcal{P}_\gamma^\text{sides}$.  The top row represents distributions classified as single-peaked while the bottom row represents distributions classified as double-peaked.  

Figure~\ref{fig: emission dist examples} shows example emission distributions from the test-particle scan, illustrating the classification algorithm. The blue curves represent the raw emission distributions from the scan, which are somewhat noisy due to the simulation's resolution of 20 time-steps per laser period. These distributions are smoothed (black curves) to determine \(\mathcal{P}_{\gamma}^\text{mid}\) (dotted-dashed line) and \(\mathcal{P}_{\gamma}^\text{sides}\) (dashed line). The ratio \(\mathcal{P}_{\gamma}^\text{mid}/\mathcal{P}_{\gamma}^\text{sides}\) is calculated for each plot, with values greater than 1 classified as single-peaked and values less than 1 classified as double-peaked. The top row displays single-peaked distributions, with decreasing \(\mathcal{P}_{\gamma}^\text{mid}/\mathcal{P}_{\gamma}^\text{sides}\) values from left to right. While larger \(\mathcal{P}_{\gamma}^\text{mid}/\mathcal{P}_{\gamma}^\text{sides}\) values correspond to more sharply peaked distributions, all examples in this row are consistently categorized as single-peaked. The bottom row shows double-peaked distributions. In panel (d), where \(\mathcal{P}_{\gamma}^\text{mid}/\mathcal{P}_{\gamma}^\text{sides} = 0\), the distribution is distinctly double-peaked. In panel (e), \(\mathcal{P}_{\gamma}^\text{mid}/\mathcal{P}_{\gamma}^\text{sides} = 0.5\), resulting in a less extreme double-peaked profile, qualitatively resembling the emission distributions in the main text from the PIC simulation with \( n_{e,0} = 4n_{cr} \) and from the test-particle calculation for the inefficient DLA example. Finally, panel (f) has \(\mathcal{P}_{\gamma}^\text{mid}/\mathcal{P}_{\gamma}^\text{sides} = 0.7\), with two side peaks and a prominent central peak. Although this distribution is more ambiguous, the algorithm classifies it as double-peaked.

To conclude this appendix, we present a plot of \(\mathcal{P}_{\gamma}^\text{mid}/\mathcal{P}_{\gamma}^\text{sides}\) for the same parameter scan shown in \cref{fig: scan} and \cref{fig: chi_ratio}. Shades of blue correspond to single-peaked distributions, while shades of green represent double-peaked distributions. Lighter green areas typically indicate borderline/ambiguous cases within the double-peaked classification. %, whereas lighter blue areas strongly align with the single-peaked category.

%To better gauge how well the algorithm classifies the number of peaks, we have Fig.~\ref{fig: grayzone} showing $\mathcal{P}_{\gamma}^\text{mid}/\mathcal{P}_{\gamma}^\text{sides}$ over the scan.  The shade of the colors quantify how strongly single-peaked or double-peaked the distribution is.  The algorithm classifies any shade of blue as single-peaked and any shade of green as double-peaked.  Informed from the examples distributions, light green distributions generally have more borderline distributions while light blue distributions tend to fit well in the single-peaked classification.

\begin{figure} [h!]
     \centering
     \includegraphics[width=0.6\linewidth]{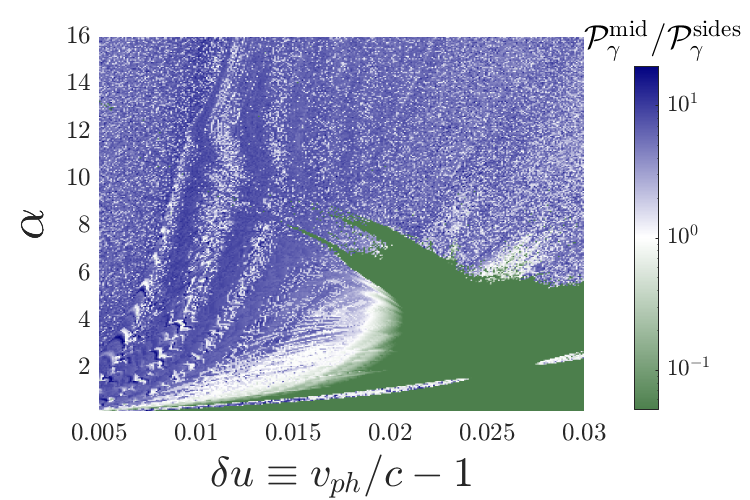}
     
     \caption{Visualization of \(\mathcal{P}_\gamma^\text{mid} / \mathcal{P}_\gamma^\text{sides}\), which compares the central and side values of the emission distribution for each electron trajectory. Large values (dark blue) indicate strongly single-peaked distributions, while small values (dark green) correspond to distinctly double-peaked distributions. Intermediate values near unity (light blue to light green) represent distributions where the classification into single- or double-peaked is less definitive. The parameter scan used to generate this plot is identical to the one shown in Fig.~12.}
     
     \label{fig: grayzone}
 \end{figure} 

%\caption{$\mathcal{P}_\gamma^\text{mid} / \mathcal{P}_\gamma^\text{sides}$ compares the central and side values of the emission distributions.  Large values (dark blue) suggest single-peaked distributions while small values (dark green) suggest double-peaked distributions.  Values near unity (light blue to light green) suggest distributions with less clear number-of-peaks determination.  The parameter scan is the same as in Fig.~\ref{fig: scan}.   }

\newpage
%%%%%%%%%%%%%%%%%%%%%%%%%%%%%%%%%%%%%%%%%%%%%%%%%%%%%%%%%%%%%%%%%%%%%%%%%%%%%%%%%%%%%%%%%%%%%%%%%%%%%%%%%%%%%%%%
\section{The Role of Radiation Friction} \label{appendix: Role RR}

This appendix examines the impact of radiation friction across a broad range of parameters by repeating the scan from \cref{table: test electron scan} over \(\alpha\) and \(\delta u\), this time excluding the force of radiation friction in Eq.~(\ref{dpdt}). In \cref{sec: test electron model}, we noted that the electron dynamics in the efficient DLA regime is strongly influenced by radiation friction. Here, we aim to assess whether this effect is a general trend, even at the moderate laser intensity considered in the main text.

\begin{figure} [h!]
     \centering
     \includegraphics[width=1.0\linewidth]{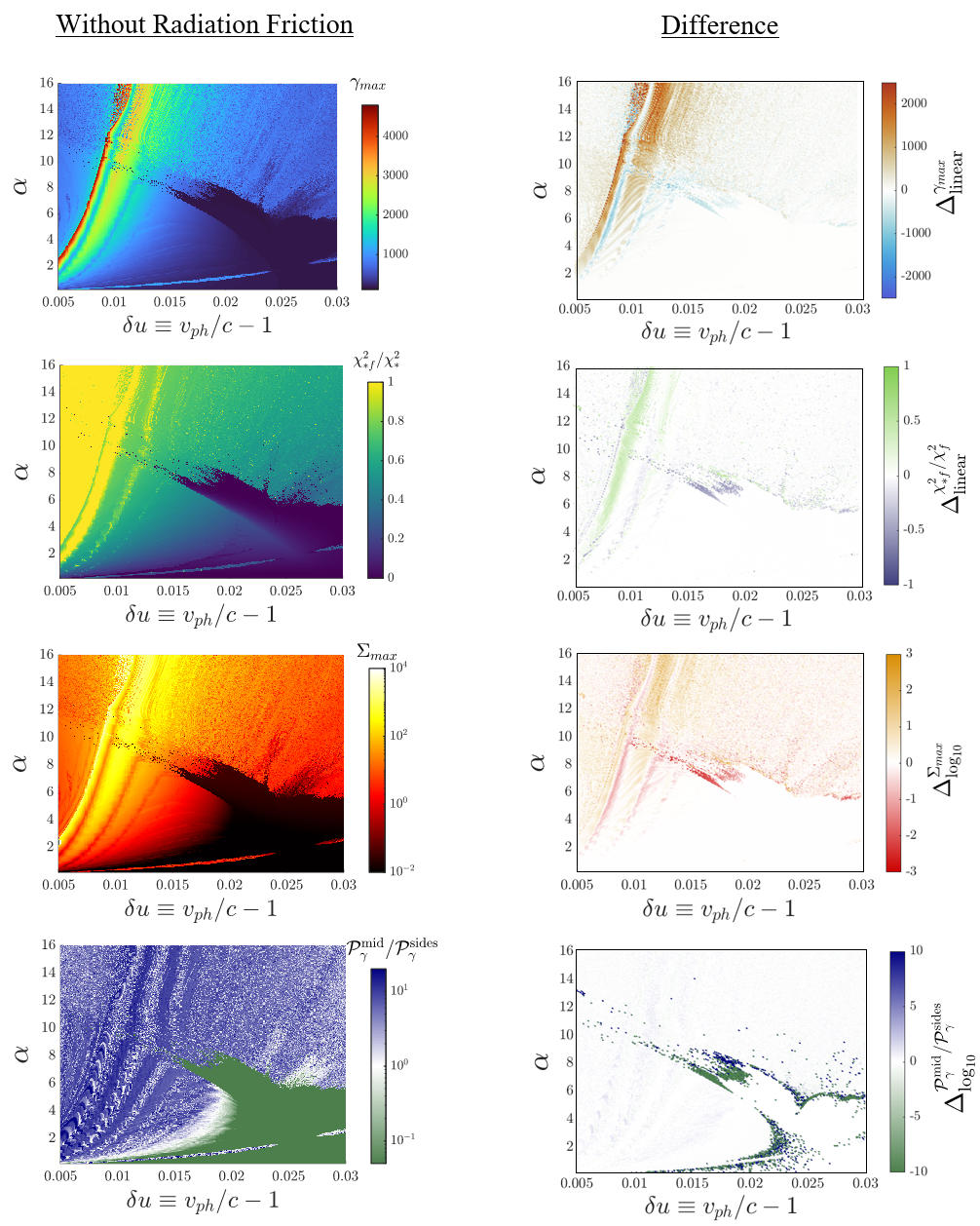}
     \caption{Comparison of electron dynamics with and without the radiation friction in the test-electron model. Left column: results of a parameter scan detailed in \cref{table: test electron scan} with the radiation friction excluded. The post-processed quantities are the same as those shown in \cref{fig: scan}, \cref{fig: chi_ratio}, \cref{fig: emission compare}, and \cref{fig: grayzone} for the scan that includes the radiation friction. Right column: the difference between the plots in the left column and the original ones. The difference is calculated using \cref{eq: diff-1} and \cref{eq: diff-2}.
     }
     \label{fig: comparison scan RR}
 \end{figure}

To compare the results of the scans with and without radiation friction, we post-processed the new scan and generated the same quantities shown in \cref{fig: scan}, \cref{fig: chi_ratio}, \cref{fig: emission compare}, and \cref{fig: grayzone}. These plots are displayed in the left column of \cref{fig: comparison scan RR}. To facilitate a comparison between these plots and those in the original figures, we calculated the differences separately, which are displayed in the right column of \cref{fig: comparison scan RR}. The method of assessing the difference depends on the dynamic range of the quantity. For quantities color-coded on a linear scale, the difference is defined as:
\begin{equation} \label{eq: diff-1}
\Delta^{\eta}_\mathrm{linear} \equiv \eta_{\sim rf} - \eta_{ rf},
\end{equation}
where \(\eta\) represents the examined quantity, and \(\eta_{\sim rf}\) and \(\eta_{ rf}\) are its values in the scans without and with radiation friction, respectively. For quantities color-coded on a logarithmic scale, the difference is defined as:
\begin{equation} \label{eq: diff-2}
\Delta^{\eta}_\mathrm{\log_{10}} \equiv \log_{10} (\eta_{\sim rf}) - \log_{10} (\eta_{ rf}).
\end{equation}

%discussed in Sec.~\ref{Sec: summary and discussion}, radiation friction significantly influences the dynamics of DLA electrons, even at moderate laser intensities such as those explored in the main text.  This is especially true in the efficient DLA regime where $\gamma$ can be many orders of magnitude greater than $a_0$ and therefore, a large quantum parameter $\chi$ is achieved resulting in impactful radiation friction.

%%%To illustrate these effects, we performed test-electron simulations excluding radiation friction.  For Fig.~\ref{fig: comparison scan RR} (left column), we scan over the same parameters as done in the plots of Sec.~\ref{sec: parameter scan} and \ref{appendix: algorithm}.  To compare this with the test-electron simulations that include radiation friction, the difference (right column) is calculated as follows:  For an arbitrary derived simulation parameter $\eta$, let $\eta_{rf}$ be the value obtained with radiation friction (main text and \ref{appendix: algorithm}) and $\eta_{\sim rf}$ be the value obtained without radiation friction (only \ref{appendix: Role RR}).  We then define 
%%%\begin{equation}
%%%\Delta^{\eta}_\mathrm{linear} \equiv \eta_{\sim rf} - \eta_{ rf}
%%%\end{equation}
%%%for linear-scale comparisons (top two rows) and 
%%%\begin{equation}
%%%   \Delta^{\eta}_\mathrm{\log_{10}} \equiv \log_{10} (\eta_{\sim rf}) - \log_{10} (\eta_{ rf})
%%%\end{equation}
%%%for log-scale comparisons (bottom two rows).

In general, the highest-energy cases show a noticeable reduction in \(\gamma_{\max}\) and \(\Sigma_{\max}\) due to radiation friction. These cases correspond to the efficient DLA regime in the scan. In contrast, the inefficient DLA regime with the lowest energies is minimally affected by radiation friction because of its relatively low emission power. Interestingly, the comparison of \(\gamma_{\max}\) reveals that removing radiation friction can result in both increases (red regions) and decreases (blue regions) in electron energy. As discussed in \cref{Sec: summary and discussion}, while radiation friction generally reduces electron energy, its interplay with DLA can also shift some electron trajectories from low- to high-energy states~\cite{Gong.SR.2019.Radiation,Yeh.NJP.2021.Friction}. Similar mixed effects are observed in the \(\rchi_{*f}^2 / \rchi_{*}^2\) ratio and \(\Sigma_{\max}\). For \(\mathcal{P}_\gamma^\text{mid} / \mathcal{P}_\gamma^\text{sides}\), which our algorithm from \ref{appendix: algorithm} uses to distinguish single- and double-peaked emission distributions, significant changes occur near the border between single- and double-peaked regions. The many orders-of-magnitude differences in \(\mathcal{P}_\gamma^\text{mid} / \mathcal{P}_\gamma^\text{sides}\) suggest that certain regions undergo dramatic transitions between efficient and inefficient DLA regimes. 

%In general, the inefficient DLA regime is minimally effected across all four parameters because of the relatively low emission power achieved.  In the $\gamma_{max}$ comparison, we see that the removal of radiation friction from the test-electron simulations can result in both increase (red) and decrease (blue) of the electron energy depending on the simulation parameters.  As mentioned in Sec.~\ref{Sec: summary and discussion}, though radiation friction generally reduces energy of the electrons, the complex interplay between DLA and radiation friction can also shift some electron trajectories from low to high energy~\cite{Gong.SR.2019.Radiation,Yeh.NJP.2021.Friction}.  We observe similarly mixed effects in the $\rchi_{*f}^2 / \rchi_{*}^2$ and $\Sigma_{max}$ from removing radiation friction.  In the $\mathcal{P}_\gamma^\text{mid} / \mathcal{P}_\gamma^\text{sides}$ difference, we see significant changes around the border of the single- and double-peaked regions.  The many orders-of-magnitude difference of $\mathcal{P}_\gamma^\text{mid} / \mathcal{P}_\gamma$ suggest certain regions transition dramatically between efficient and inefficient DLA regimes.  

%In the left column are simulation scans ($\gamma_{max}$, $\rchi_{*f}^2 / \rchi_{*}^2$, $\Sigma_{max}$, $\mathcal{P}_\gamma^\text{mid} / \mathcal{P}_\gamma^\text{sides}$) with the same parameters as detailed in Table~\ref{table: test electron scan} but where test-electrons do not experience radiation friction.  On the right are the differences ($\Delta^{\gamma_{max}}_\mathrm{linear}$, $\Delta^{\rchi_{*f}^2 / \rchi_{*}^2}_\mathrm{linear}$, $\Delta^{\Sigma_{max}}_\mathrm{log_{10}}$ , $\Delta^{\mathcal{P}_\gamma^\text{mid} / \mathcal{P}_\gamma^\text{sides}}_\mathrm{log_{10}}$) comparing these scans with the previous version where test-electrons do experience radiation friction.

\newpage

%%%%%%%%%%%%%%%%%%%%%%%%%%%%%%%%%%%%%%%%%%%%%%%%%%%%%%%%%%%%%%%%%%%%%%%%%%%%%%%%%%%%%%%%%%%%%%%%%%%%%%%%%%%%%%%%
\end{document}